\documentclass[twocolumn,aps,prl,superscriptaddress,longbibliography]{revtex4-2}
\usepackage{graphicx}
\usepackage{amsmath}
\usepackage{amssymb}
\usepackage{braket}
\usepackage{physics}
\usepackage[colorlinks=true,linkcolor=blue,citecolor=blue,urlcolor=blue]{hyperref}

\begin{document}

\title{Exceptionally Slow Relaxation from Micro-canonical to Canonical Ensembles in Quasi-one-dimensional Quantum Gases}
\author{Huaichuan Wang}
\thanks{These authors contributed equally.}
\affiliation{Department of Physics and State Key Laboratory of Low Dimensional Quantum Physics, Tsinghua University, Beijing, 100084, China}
\author{Xixiang Du}
\thanks{These authors contributed equally.}
\affiliation{Institute for Advanced Study, Tsinghua University, Beijing 100084, China}
\author{Zhongchi Zhang}
\thanks{These authors contributed equally.}
\affiliation{Department of Physics and State Key Laboratory of Low Dimensional Quantum Physics, Tsinghua University, Beijing, 100084, China}
\author{Yue Wu}
\affiliation{Institute for Advanced Study, Tsinghua University, Beijing 100084, China}
\author{Ken Deng}
\affiliation{Department of Physics and State Key Laboratory of Low Dimensional Quantum Physics, Tsinghua University, Beijing, 100084, China}
\author{Zihan Zhao}
\affiliation{Quantum Science Center of Guangdong-Hong Kong-Macao Greater Bay Area, Shenzhen, 518045, Guangdong, China}
\author{Chengshu Li}
\affiliation{Institute for Advanced Study, Tsinghua University, Beijing 100084, China}
\author{Zheyu Shi}
\affiliation{State Key Laboratory of Precision Spectroscopy, East China Normal University, Shanghai 200062, China}
\author{Wenlan Chen}
\email{cwlaser@ultracold.cn}
\affiliation{Department of Physics and State Key Laboratory of Low Dimensional Quantum Physics, Tsinghua University, Beijing, 100084, China}
\author{Hui Zhai}
\email{hzhai@tsinghua.edu.cn}
\affiliation{Institute for Advanced Study, Tsinghua University, Beijing 100084, China}
\affiliation{Hefei National Laboratory, Hefei 230088, China}
\author{Jiazhong Hu}
\email{hujiazhong01@ultracold.cn}
\affiliation{Quantum Science Center of Guangdong-Hong Kong-Macao Greater Bay Area, Shenzhen, 518045, Guangdong, China}
\date{\today}

\begin{abstract}

Integrability in one dimension prevents quantum thermalization and gives rise to rich many-body phenomena described by generalized hydrodynamics, which have been extensively studied over the past two decades using cold atoms in optically confined tubes. However, experimental work to date has focused primarily on low-energy states. Here, we report the experimental observation and theoretical understanding of near-integrable effects on thermalization in highly excited states. We design a protocol to prepare atoms within a high-energy window by combining a harmonic trap and a weak optical lattice: a Bose-Einstein condensate is initially prepared away from the trap center via Wannier-Stark localization and subsequently emits atoms into a selected energy window of highly excited states via Landau-Zener tunneling. By reconstructing the Wigner functions from the density distribution using a machine learning algorithm, we find that it takes an exceptionally long time, up to several seconds, for these atoms to gradually thermalize from an approximately microcanonical ensemble toward a canonical ensemble. We develop a modified Boltzmann equation that captures weak integrability breaking, yielding good agreement between theory and experiment. Our results extend the understanding of integrability and thermalization in low-dimensional quantum systems.

\end{abstract}

\maketitle

Quantum thermalization in isolated systems underpins quantum statistical mechanics \cite{Deutsch1991ClosedSystem, Srednicki1994ChaosThermalization, Rigol2008ThermalizationMechanism, Dalessio2016ETHReview}. Given the ubiquity of thermalization, research in this field often focuses on mechanisms that can prevent or violate it, among which integrability plays a central role \cite{Rigol2009BreakdownThermalization, Polkovnikov2011ColloquiumNonequilibrium, Mori2018ThermalizationPrethermalization}. In one dimension, when two particles undergo an elastic collision with conserved total momentum and energy, only two outcomes are possible: they either retain their original momenta or exchange them. This unique dynamical constraint forms the microscopic basis for integrability in one-dimensional systems and gives rise to the well-known Newton's cradle effect. Ultracold atoms confined in quasi-one-dimensional tubes by optical lattices provide an ideal platform for studying integrable quantum many-body systems. Indeed, a quantum Newton's cradle was realized in such a setting two decades ago \cite{Kinoshita2006QuantumNewtonsCradle}. This landmark experiment sparked extensive research, leading to the emergence of the theoretical frameworks known as generalized hydrodynamics and generalized Gibbs ensemble \cite{Rigol2007RelaxationIntegrable, Vidmar2016GGEIntegrableLattices, CastroAlvaredo2016EmergentHydrodynamics, Bertini2016TransportXXZ}, which have since been used to uncover a wealth of richer phenomena both theoretically \cite{Cassidy2011GeneralizedThermalization, Kollar2011GGEPrethermalization, Ilievski2015CompleteGGE, Bulchandani2017SolvableHydrodynamics, DeNardis2018HydrodynamicDiffusion, DeNardis2019DiffusionGHD} and experimentally \cite{Gring2012RelaxationPrethermalization, Langen2015ExperimentalGGE, Schemmer2019GHDAtomChip, Malvania2021GHDStronglyInteracting, Moller2021ExtensionGHD, Cataldini2022EmergentPauliBlocking, Rosenberg2024DynamicsMagnetization}.

However, these efforts have predominantly focused on low-energy dynamical processes. Here, we consider the opposite limit, in which the atomic kinetic energy far exceeds all other energy scales and inter-particle interactions are relatively weak. In this regime, the many-body system can be effectively described by a statistical ensemble of single particles. Suppose we prepare an initial state in which the system approximately resides in a microcanonical ensemble, that is, all atoms possess nearly identical energy, or more precisely, their energies are confined to a narrow window. Then, owing to the constrained kinematics in one dimension, the energy distribution remains unchanged even after elastic collisions. Consequently, the system will persist in a state close to a microcanonical ensemble for a sufficiently long time despite ongoing collisions. This behavior can be viewed as a high-energy manifestation of Newton's cradle, a phenomenon that has not been observed experimentally until now.

\begin{figure*} 
	\centering 
	\includegraphics[width=0.9\textwidth]{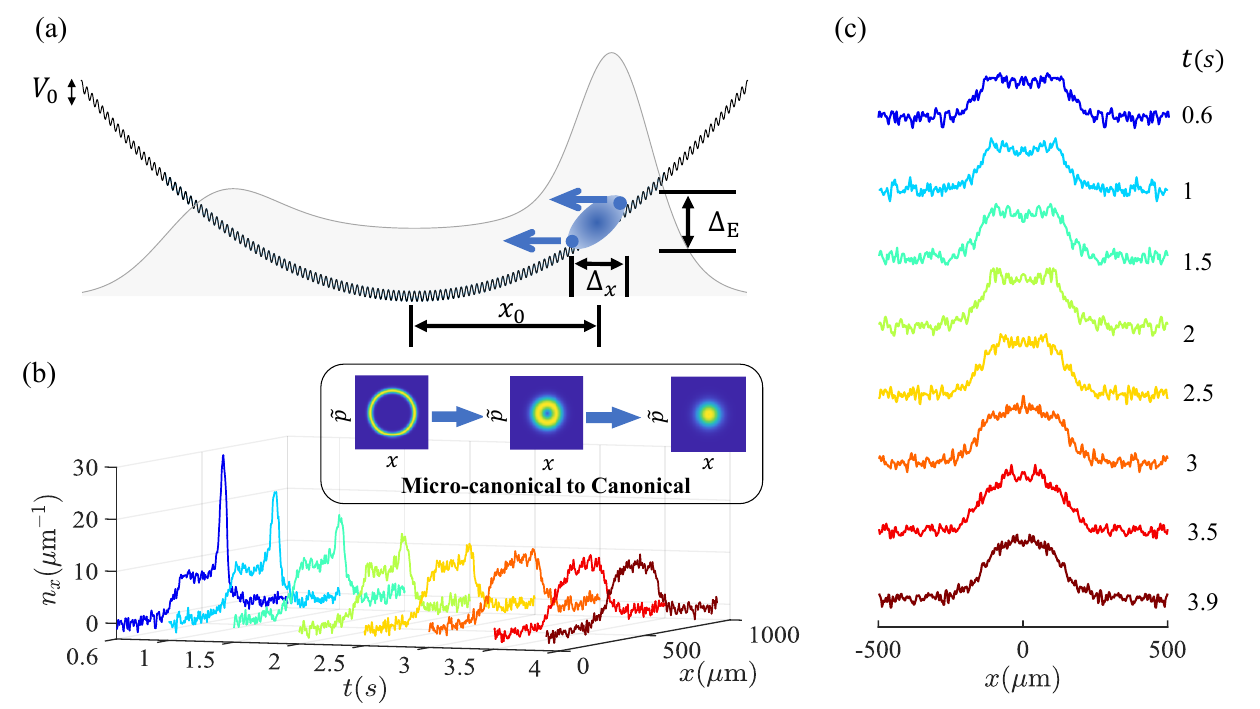} 
	\caption{(a) Experimental setup. A 1560-nm laser creates a quasi-one-dimensional harmonic trap. A BEC, initially prepared in a 1064-nm dipole trap, is displaced by $x_0=120\mathrm{\mu m}$ from the center of the confining potential. Simultaneously, the BEC is subjected to a shallow optical lattice with depth $V_0 = 0.375 E_r$. The BEC is localized via Wannier-Stark effects and continuously emits high-energy delocalized atoms through Landau-Zener tunneling. The gray shadow indicates the total density resulting from both the delocalized atoms and the condensate. (b) Evolution of the atomic density profile. A plateau-like density profile forms near the center of the harmonic trap and persists throughout the subsequent evolution up to $4\text{s}$. It eventually thermalizes into a Gaussian distribution due to the breakdown of integrability. The inset illustrates a schematic evolution of the Wigner functions: from a large and narrow ring representing a steady state of an approximately microcanonical ensemble to a final Gaussian, corresponding to a steady state of the canonical ensemble. (c) Evolution of the delocalized atoms. Using the waist position of the 1560-nm trap as a reference, we retain only atoms on its left side and symmetrize the distribution. A clear plateau with a central dip is observed. The plateau gradually evolves into a Gaussian distribution at sufficiently long times.} 
		\label{illustration} 
\end{figure*}
			
The challenge in observing this phenomenon lies in preparing a high-energy microcanonical ensemble. To address this, we design a specific experimental protocol, as illustrated in Fig.~\ref{illustration}(a). First, we create a quasi-one-dimensional harmonic trap using a 1560-nm laser, with an axial trapping frequency of $\omega_{\parallel}=2\pi\times6.7\text{Hz}$, a beam waist of $25\mathrm{\mu m}$, and a radial confinement frequency of $\omega_{\perp}=2\pi\times514.3\text{Hz}$. Additionally, a weak lattice with depth $V_0=0.375E_\text{r}$ is superimposed along the axial direction, where $E_\text{r}=h\times972.6\text{Hz}$. A Bose-Einstein condensate (BEC) of $^{85}Rb$ is initially produced at a position displaced by $x_0=120\mathrm{\mu m}$ from the center of the harmonic trap, confined by a separate tight trap generated with a 1064-nm laser.

Upon turning off the 1064-nm laser confinement, for a sufficiently large displacement $x_0$, the resulting gradient force becomes strong enough that the condensate undergoes Bloch oscillations with a very small amplitude. This effectively localizes the BEC near its initial position, a manifestation of Wannier-Stark localization \cite{Emin1987WannierStark, Morsch2001BlochOscillations}. Concurrently, Landau-Zener tunneling induces interband transitions, emitting delocalized atoms \cite{Zhao2026StarkMBLContinuous}. As these atoms move toward the trap center, they acquire substantial kinetic energy, $E_0= \frac{1}{2} m \omega_\parallel^2 x_0^2=h\times 2715\text{Hz}$, owing to the large initial displacement $x_0$ \cite{supple}. The tight confinement of the 1064-nm laser, together with interatomic interactions, sets the width $\Delta_x$ of the condensate, which in turn determines the energy spread $\Delta_E = m \omega_\parallel^2 x_0 \Delta_x$. The condition $\Delta_x \ll x_0$ ensures $\Delta_E \ll E_0$. In this manner, the delocalized atoms emitted from the BEC constitute an approximately microcanonical ensemble.

\begin{figure*} 
	\centering 
	\includegraphics[width=0.9\textwidth]{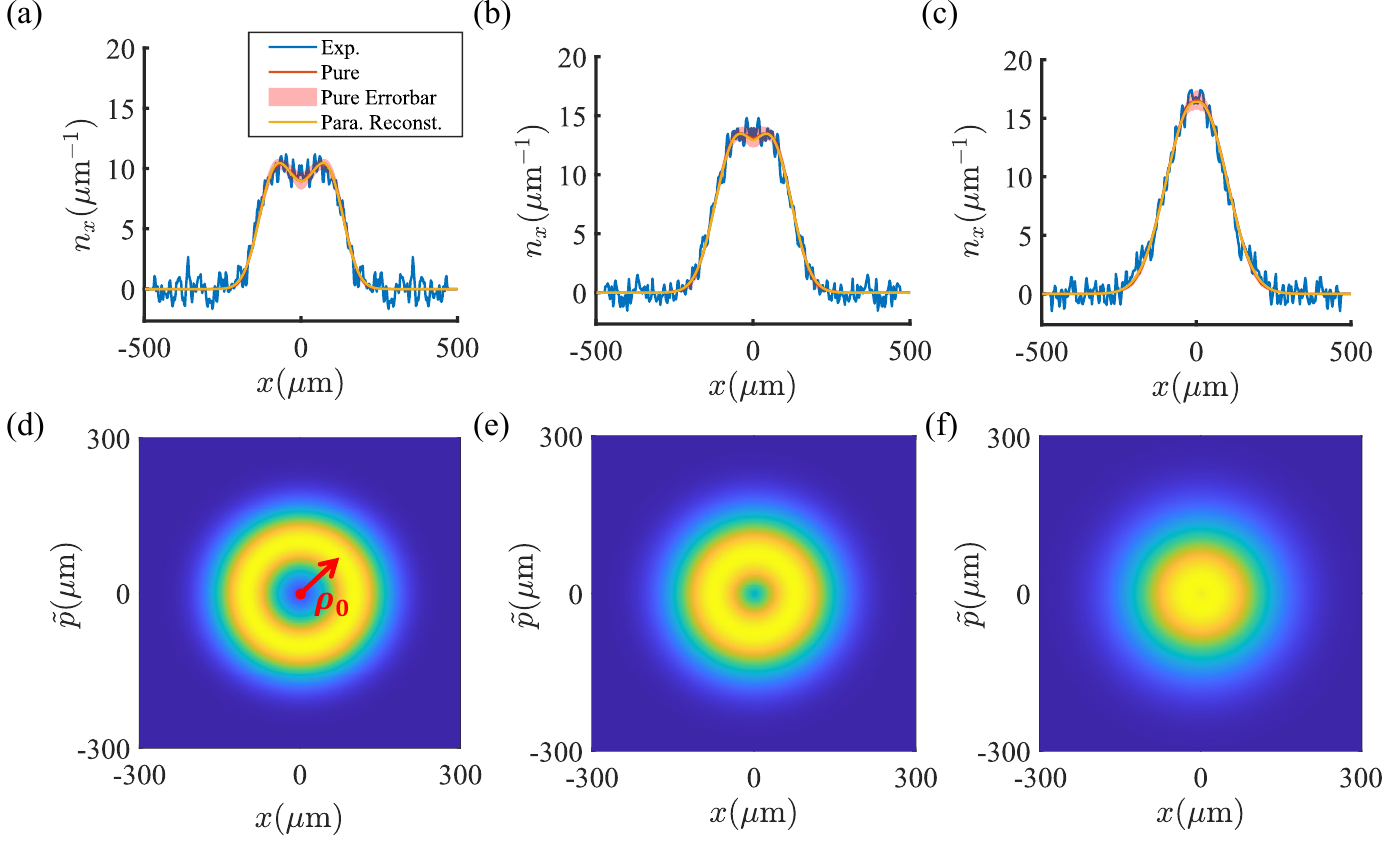} 
	\caption{Reconstruction of the Wigner functions. (a)-(c) show three typical density distributions and (d)-(f) show their corresponding reconstructed Wigner functions. In the top row, the blue curves represent the symmetrized density profile; the red curves and the red shaded areas denote the denoised data and the uncertainty band, respectively. The bottom row displays the reconstructed Wigner functions $f(x,\tilde{p})$. The yellow curves in the top row plot the density profiles obtained from $n(x)=\int f(x,\tilde{p}) \mathrm{d}\tilde{p}$.} 
		\label{Wigner} 
\end{figure*}

Therefore, the density profile of these delocalized atoms resembles that of a high-energy excited state in a harmonic trap, with a flat plateau and a slight dip at the center. The total density profile comprises both the delocalized atoms and the condensed atoms, as indicated by the gray shaded region in Fig.~ \ref{illustration}(a). A typical time evolution of the density profile is shown in Fig.~\ref{illustration}(b). It can be seen that the BEC portion gradually loses atoms due to continuous emission via Landau-Zener tunneling, while the delocalized atoms exhibit plateau-like behavior and slowly evolve toward a Gaussian distribution. To better focus on the density of the delocalized atoms, we note that the BEC remains localized on the right side, far away from the trap center. It is reasonable to assume that atoms on the left side of the trap are purely delocalized and that their density profile should be symmetric. Thus, we retain only the density on the left side and symmetrize it about the trap center, yielding the time evolution of the density profile for the delocalized atoms alone, as shown in Fig. ~\ref{illustration}(c). 
		
The real-space density $n(x)$ is obtained from the Wigner function in phase space $f(x,\tilde{p})$ via the relation $n(x)=\int \mathrm{d}\tilde{p} f(x,\tilde{p})$, where $\tilde{p}=p/(m\omega_{\parallel})$ is the normalized momentum along $\hat{x}$. Reconstructing the Wigner functions $f(x,\tilde{p})$ from the measured real-space density profiles, as illustrated in Fig.~\ref{Wigner}(d)-(f), can provide valuable insight. This reconstruction, however, poses a challenging inverse problem, exacerbated by experimental noise in the density data. To address this, we employ a machine-learning approach. 

For our system, we assume the Wigner functions of the dataset for model training follow the ansatz
\begin{equation}
f(x,\tilde{p}) = A \exp\left\{-\frac{(\rho-\rho_0)^2}{2\sigma_{\rho}^2}\right\}, \label{wigner_uniform}
\end{equation}
where $\rho^2 = x^2 + \tilde{p}^{2}$, and a constant $\rho$ corresponds to a constant single-particle energy in the axial harmonic trap. Here, we ignore the presence of a weak lattice along the axial direction because the mean energy of these delocalized atoms $E_0$ is so high compared to the lattice depth $V_0$ that the lattice modification of single-particle dispersion is completely negligible. The parameter $A$ is a normalization constant ensuring unit total probability. In the limit $\rho_0 \gg \sigma_\rho$, the Wigner function is confined to a narrow ring, corresponding to nearly fixed single-particle energies, and the limit $\sigma_\rho \rightarrow 0$ recovers the microcanonical ensemble. Conversely, in the limit $\sigma_\rho \gg \rho_0$, the Wigner function approaches a broad Gaussian distribution, and the limit $\rho_0 \rightarrow 0$ recovers the canonical ensemble.

To train the reconstruction algorithm, we first measure the background noise distribution $\delta n(x)$ in the absence of atoms. We then generate a training dataset by constructing input signals $n(x)+\delta n(x)$ for various choices of $\rho_0$, $\sigma_\rho$, and random noise instances $\delta n(x)$. The corresponding output labels are the parameters ${\rho_0, \sigma_\rho}$ that characterize the Wigner functions. A supervised learning protocol is employed to reduce noise and extract these parameters \cite{supple}. Once trained, the algorithm is applied directly to the symmetrized experimental density profiles of the delocalized atoms to reconstruct their underlying Wigner functions. Typical results are shown in Fig.~\ref{Wigner}. In Fig.~\ref{Wigner}(a)-(c), the red lines with error bars represent the noise-reduced density profiles, while the yellow lines present the density profiles generated by the corresponding Wigner functions shown in Fig.~\ref{Wigner}(d)-(f). The excellent overlap of these curves supports the reliability of the extracted Wigner functions.

\begin{figure} 
\includegraphics[width=0.48\textwidth]{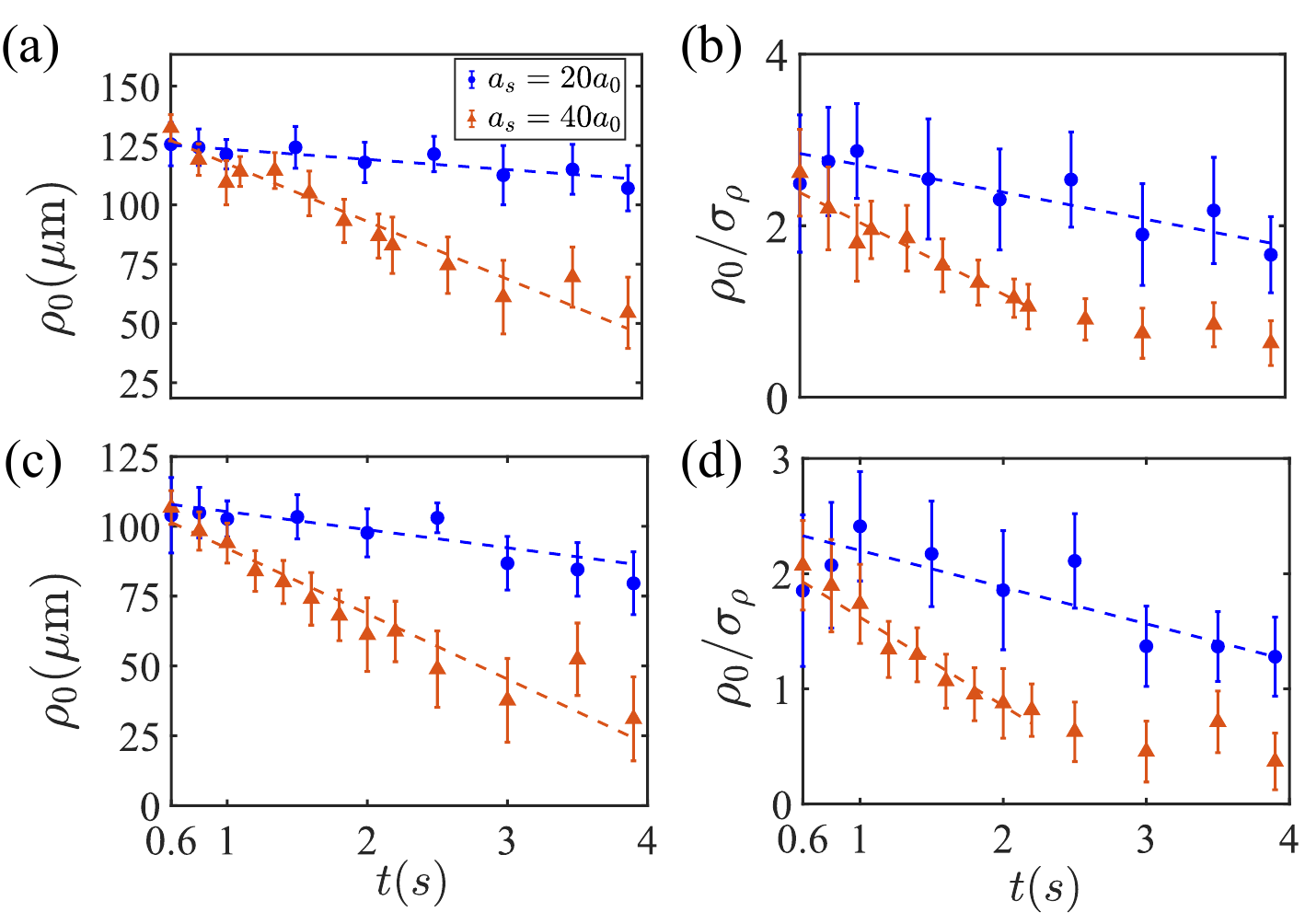} 
	\caption{The time evolution of $\rho_0$ (panels a, c) and $\rho_0/\sigma_\rho$ (panels b, d) is shown for two interaction strengths, $20a_0$ (blue circles) and $40a_0$ (orange triangles), for initial spatial widths of BEC $\Delta_x=7\mathrm{\mu m}$ (a, b) and $\Delta_x = 10\mathrm{\mu m}$ (c, d), respectively. Here $a_0$ is the Bohr radius. The dashed lines represent linear fits to the data. }
		\label{thermalization} 
\end{figure}

Figure~\ref{thermalization} presents the time evolution of $\rho_0$ and $\rho_0/\sigma_\rho$ for different scattering lengths $a_\text{s}$ and different initial spatial widths $\Delta_x$ of the BEC. Here, the initial value of $\rho_0$ is determined by the mean initial energy $E_0$, and the initial $\sigma_\rho$ reflects the initial energy spreading fixed by $\Delta_E$. The linear decrease in $\rho_0$ and $\rho_0/\sigma_\rho$ over time signals a smooth crossover from an approximately microcanonical ensemble to a canonical ensemble. Notably, this crossover occurs over several seconds, a timescale substantially longer than the typical timescales of processes in ultracold atomic systems.  

This exceptionally long microcanonical-to-canonical relaxation time arises directly from the integrability and its weak breaking in one dimension. Given that these delocalized atoms are in highly excited states, their dynamics are more suitably described by the Boltzmann equation:
\begin{equation}
\frac{\partial f}{\partial t} + \frac{p}{m}\frac{\partial f}{\partial x} + F(x)\frac{\partial f}{\partial p} = I_{\mathrm{coll}}[f].
\label{EQ_boltzman}
\end{equation}
Here, $F(x)$ is the force from the background harmonic trap. The collision integral $I_{\mathrm{coll}}[f]$, which accounts for two-body collision events, is given by its standard form
\begin{equation}
I_{\mathrm{coll}}[f_1] = \int \mathrm{d}p_2 \frac{|p_1-p_2|}{m} \big[ f^\prime_1 f^\prime_2 - f_1 f_2 \big],
\label{EQ_collision}
\end{equation}
where $f_i \equiv f(x,p_i,t)$ and $f_i' \equiv f(x,p_i',t)$ are the Wigner functions evaluated at the incoming momenta of the direct and inverse collisions. The term $f'_1 f'_2$ describes collisions from momenta $(p'_1,p'_2)$ resulting in $(p_1,p_2)$, while $f_1 f_2$ describes the corresponding collisions from $(p_1,p_2)$ into the allowed outgoing momenta. In one dimension, the only possible outcomes are either $(p^\prime_1,p^\prime_2) = (p_1,p_2)$ or an exchange of the two momenta; both result in $f^\prime_1 f^\prime_2 = f_1 f_2$ and therefore $I_{\mathrm{coll}}[f_1]=0$. Consequently, the Wigner functions do not evolve.

Therefore, to explain the observed dynamics, a weak integrability-breaking term must be introduced into the Boltzmann equation. Here, we choose to relax strict energy conservation by replacing it with
\begin{equation}
\frac{p^2_1}{2m}+\frac{p^2_2}{2m}+\epsilon=\frac{p^{\prime 2}_1}{2m}+\frac{p^{\prime 2}_2}{2m},
\label{Energy_conservation_relaxing}
\end{equation}
where $p_{1,2}$ and $p'_{1,2}$ are the momenta before and after collision. This reflects that the axial energy of a colliding pair is not strictly conserved in this quasi-one-dimensional system. The mismatch $\epsilon$ corresponds to energy transfer into transverse modes, a process governed by a probabilistic function $g(\epsilon)$. Accordingly, we propose a modified phenomenological collision integral
\begin{equation}
I_{\mathrm{coll}}[f_1] = \int g(\epsilon)\mathrm{d}\epsilon \int \mathrm{d}p_2 \frac{|p_1-p_2|}{m} \big[ f^\prime_1 f^\prime_2 - f_1 f_2 \big],
\label{EQ_collision_mod}
\end{equation}
which is generally nonzero and can drive the time evolution of the Wigner functions. Nevertheless, the Boltzmann distribution should remain an equilibrium steady state; that is, $I_{\mathrm{coll}}[f]$ must vanish when all four $f_i$ and $f^\prime_i$ follow the Boltzmann distribution. 

\begin{figure} 
\includegraphics[width=0.45\textwidth]{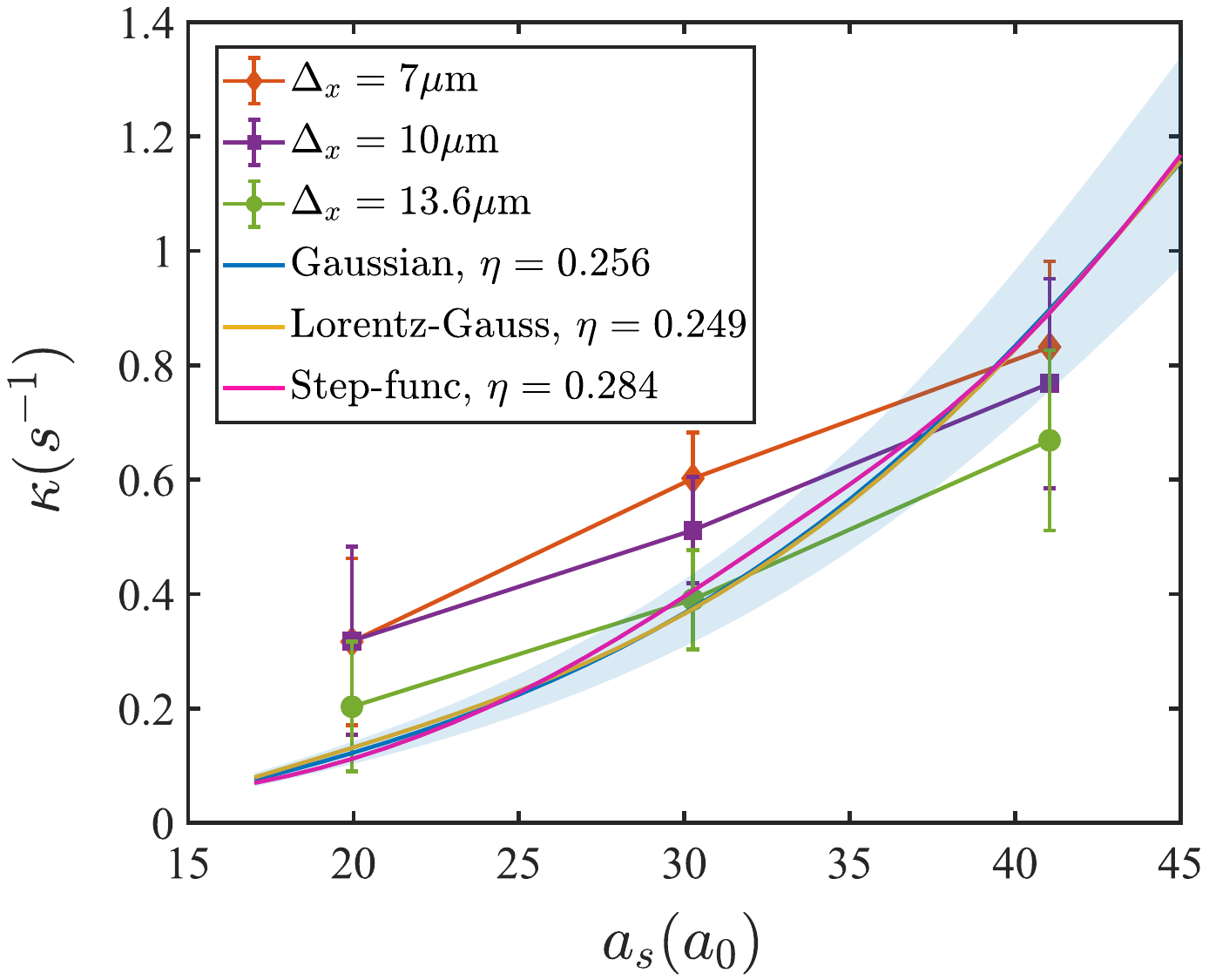} 
	\caption{The thermalization rate $\kappa$ as a function of the scattering length $a_\text{s}$ for different initial widths $\Delta_x$. The blue line shows the rate obtained from numerical simulation with the Gaussian ansatz. The light blue band indicates the standard error due to uncertainty in the atom number. The yellow and pink lines show the rate obtained from numerical simulation with the Lorentzian-Gaussian ansatz and step-function-like ansatz respectively.}
		\label{exp-theory} 
\end{figure}
		
This condition is satisfied provided that $g(\epsilon)$ takes the form $g(\epsilon)=e^{-\frac{\beta\epsilon}{2}}h(\epsilon)$, where $h(\epsilon)$ must be an even function \cite{supple}. Here $\beta$ is the inverse temperature of the final canonical ensemble. The function $h(\epsilon)$ should decay sufficiently rapidly so that $g(\epsilon)$ is normalizable and its first moment is finite. Because of the Boltzmann factor $e^{-\beta\epsilon/2}$, positive and negative energy mismatches are weighted asymmetrically, implying a net transfer of energy from the axial degree of freedom to transverse modes during relaxation, consistent with the experimentally observed contraction of the ring-like Wigner functions. Importantly, the relaxation rate is controlled by the standard deviation $\sigma$ of $h(\epsilon)$. Hence, we can directly use $\sigma$ to parametrize the function $h$ as $h(\epsilon/\sigma)$.
		
Figure~\ref{exp-theory} shows the relaxation rate $\kappa$, extracted from the linear change of $\rho_0/\sigma_\rho$ over time, as a function of the scattering length $a_\text{s}$. In the measurements, we adjust the frequency of the 1064-nm confinement laser simultaneously with $a_\text{s}$ to keep $\Delta_x$ fixed. For a given $\Delta_x$, $\kappa$ increases monotonically with $a_\text{s}$. This is expected, as a larger interaction energy allows more energy to be transferred to transverse modes during a collision. We therefore assume $\sigma = 4\pi\eta \hbar^2 a_\text{s}/(ma^3_\perp)$, with $a_\perp=\sqrt{\hbar/(m\omega_\perp)}$ being the transverse harmonic length, and estimate $\eta$ by solving the two-body problem in a quasi-one-dimensional geometry \cite{Olshanii1998ConfinementInduced, Bergeman2003CIR, Moore2004TightWaveguides}, which yields $\eta \approx 0.567$ \cite{supple}.

We also numerically solve the modified Boltzmann equation to obtain $\kappa$ as a function of $\sigma$ \cite{supple}. To fairly compare theory with experiment, we consider three different forms of $h(\epsilon)$ characterized by the same standard deviation $\sigma$: (i) the Gaussian ansatz $h(\epsilon) \propto \exp[-\epsilon^2/(2\sigma^2)]$, (ii) the Lorentzian-Gaussian ansatz $h(\epsilon) \propto (\epsilon^2 + w^2)^{-1} \exp[-\epsilon^2/(2w^2)]$ with $w = 1.38\sigma$, and (iii) the step-function-like ansatz $h(\epsilon) \propto \Theta(\sqrt{3}\sigma - |\epsilon|)$. In each case, $\eta$ is treated as a fitting parameter to achieve the best agreement between theory and experiment. This procedure yields $\eta = 0.256$ (Gaussian), $0.249$ (Lorentzian-Gaussian), and $0.284$ (step-function), all of which are of the same order of magnitude with the estimate from the microscopic calculation. Moreover, the theoretical results shown in Fig.~\ref{exp-theory} indicate that the relaxation dynamics is not sensitive to the specific form chosen for $h(\epsilon)$, provided that its standard deviation is fixed. These results support that our modified Boltzmann equation provides a proper microscopic description of the slow microcanonical-to-canonical relaxation.

In summary, we observe a slow relaxation from an approximately microcanonical ensemble toward the canonical ensemble, a novel dynamical effect in a near-integrable one-dimensional system. This takes advantage of our particular scheme for preparing an ensemble of thermal atoms within a narrow high-energy window. We attribute the main mechanism for breaking integrability in such a high-energy system to energy transfer to the transverse degree of freedom, and we propose a modified Boltzmann equation to describe the relaxation process, achieving reasonable agreement with experiment. Our results complement previous investigations of near-integrable thermalization dynamics in low-energy states.		

\textit{Acknowledgments.} This work is supported by National Key Research and Development Program of China under Grant No. 2021YFA0718303 (J.H.), No. 2021YFA1400904 (W.C.) and No. 2023YFA1406702 (W.C. and H.Z.); and National Natural Science Foundation of China under Grant No. 92476110 (J.H.), No. 92576208 (W.C.), No. 12488301 (H.Z.) and No. U23A6004 (H.Z.).
		
\bibliography{reference}

\end{document}


\title{Supplementary\\Exceptionally Slow Relaxation from Micro-canonical to Canonical Ensembles in Quasi-one-dimensional Quantum Gases}

\maketitle

\section{Fast dephasing induced by a shallow lattice}
\subsection{Time Evolution of the dephasing process}
\begin{figure}[htbp]
    \centering
    \includegraphics[width=0.9\linewidth]{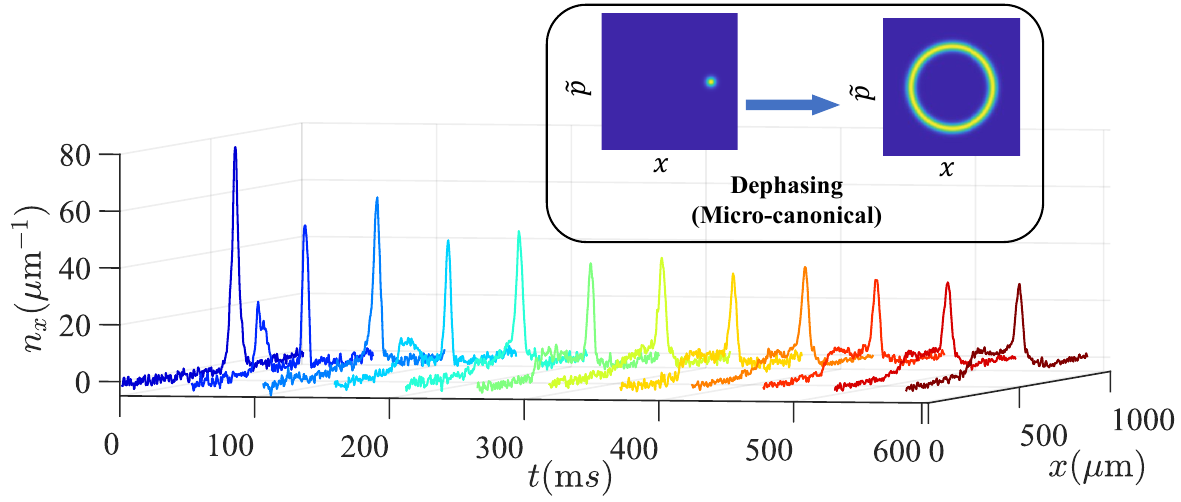}
    \caption{Time evolution of dephasing. After sliding down the trap, the delocalized atoms oscillate in the trap. The presence of the shallow lattice induces rapid dephasing. After about $600\,$ms, the dephased atoms develop a characteristic plateau with a central dip in its density profile. The inset shows a schematic diagram of the Wigner function: from a Gaussian packet to an angular-uniform ring, which can be regarded as a steady state of the approximately microcanonical ensemble.}
    \label{FigS:First_Timescale}
\end{figure}

To prepare a high-energy microcanonical ensemble, the delocalized atoms should dephase as fast as possible. We use a shallow lattice to accelerate this process. In our experiment, we observe dephasing that occurs significantly faster than what would be expected from trap anharmonicity alone. The evolution is shown in Fig.~\ref{FigS:First_Timescale}. We observe that sliding atoms oscillate in the trap at first. After around four periods of evolution($600\,$ms), the oscillation rapidly decays and forms a plateau-like density profile with a central dip. This process can be described in the phase-space. At first, the Wigner function is a concentrated Gaussian packet rotating along a ring whose radius is determined by the initial atomic position $x_0$. Then the dephasing process spreads the Wigner function uniformly over the constant-energy ring from a concentrated packet. And such a circular ring with uniform angular distribution in phase-space will give rise to this density profile. 

We attribute this enhanced dephasing to the presence of a shallow lattice. The background trap is created by a single 1560-nm beam. Approximately $0.25\%$ of the laser power is reflected, producing a lattice with a depth of only $0.375\,E_r$. This shallow lattice plays a crucial role in the rapid dephasing observed during the first stage of the dynamics. 

\subsection{Anharmonicity of Quasi-1D Trap}
\begin{figure}
    \centering
    \includegraphics[width=0.8\linewidth]{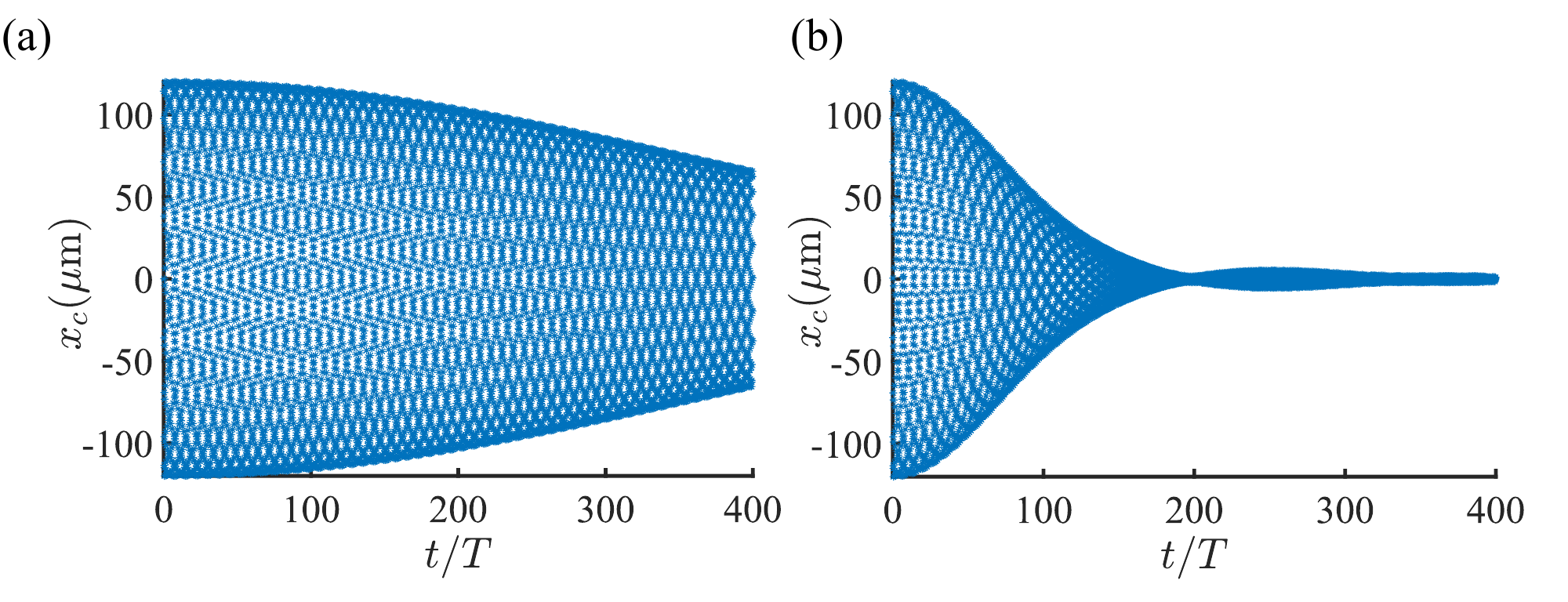}
    \caption{Time evolution of the center of mass under trap anharmonicity. Here $T=2\pi/\omega_z$ is the oscillation period. (a) For an initial Gaussian width of $\sigma_x=4\mathrm{\mu m}$(corresponding to a Thomas–Fermi radius of $\Delta_x=14\,\mathrm{\mu m}$), the atoms remain coherent after 400 oscillation periods, showing no appreciable dephasing. (b) Even when the initial width is increased to  $\sigma_x=20\,\mathrm{\mu m}$, roughly 150 periods are required for the ensemble to dephase.}
    \label{FigS1:ball_dephasing}
\end{figure}

For a 1560-nm laser beam with a waist of $25\,\mathrm{\mu m}$, the resulting trapping potential takes the form
\begin{equation}
    V(z)=V_0\frac{1}{1+(z/z_R)^2}
    \label{anharmonic_equation}
\end{equation}
where $z_R=\pi w_0^2/\lambda$ is the Rayleigh range, $V_0=\omega_z^2 m \pi^2 w_0^4/(2\lambda^2)$, $\omega_z$ is the axial harmonic frequency, $m$ is the mass of $^{85}Rb$, $w_0$ is the waist of laser beam and $\lambda$ is the wavelength of the laser. We treat the atomic ensemble as a set of noninteracting classical particles moving in the potential described by Eq.~\ref{anharmonic_equation}. We simulate 1000 classical particles initially placed at a distance $x_0=120\,\mathrm{\mu m}$ from the trap center with a Gaussian spatial width $\sigma_x$ and assume zero initial velocity for all particles. After computing the trajectory of each particle, we evaluate the center-of-mass position $x_c$ of the ensemble at each time. Once dephasing induced by the trap anharmonicity is complete, the center of mass relaxes to zero. Our simulations show that, for parameters relevant to the experiment, anharmonicity alone cannot lead to the observed early-time dephasing. For an initial Gaussian width of $\sigma_x=4\,\mathrm{\mu m}$ (corresponding to a Thomas–Fermi radius of $\Delta_x=14\,\mathrm{\mu m}$), the center-of-mass oscillation remains coherent even after 400 oscillation periods; see Fig.~\ref{FigS1:ball_dephasing}(a). Even when the width is increased to $\sigma_x=20\,\mathrm{\mu m}$, far exceeding the experimental value, roughly 150 periods are required for substantial dephasing; see Fig.~\ref{FigS1:ball_dephasing}(b). These results indicate that the first-stage dephasing observed in the experiment cannot be attributed to the weak anharmonicity of the trap.

\subsection{Motion in the Band Structure Induced by Shallow Lattice}
\begin{figure}
    \centering
    \includegraphics[width=0.9\linewidth]{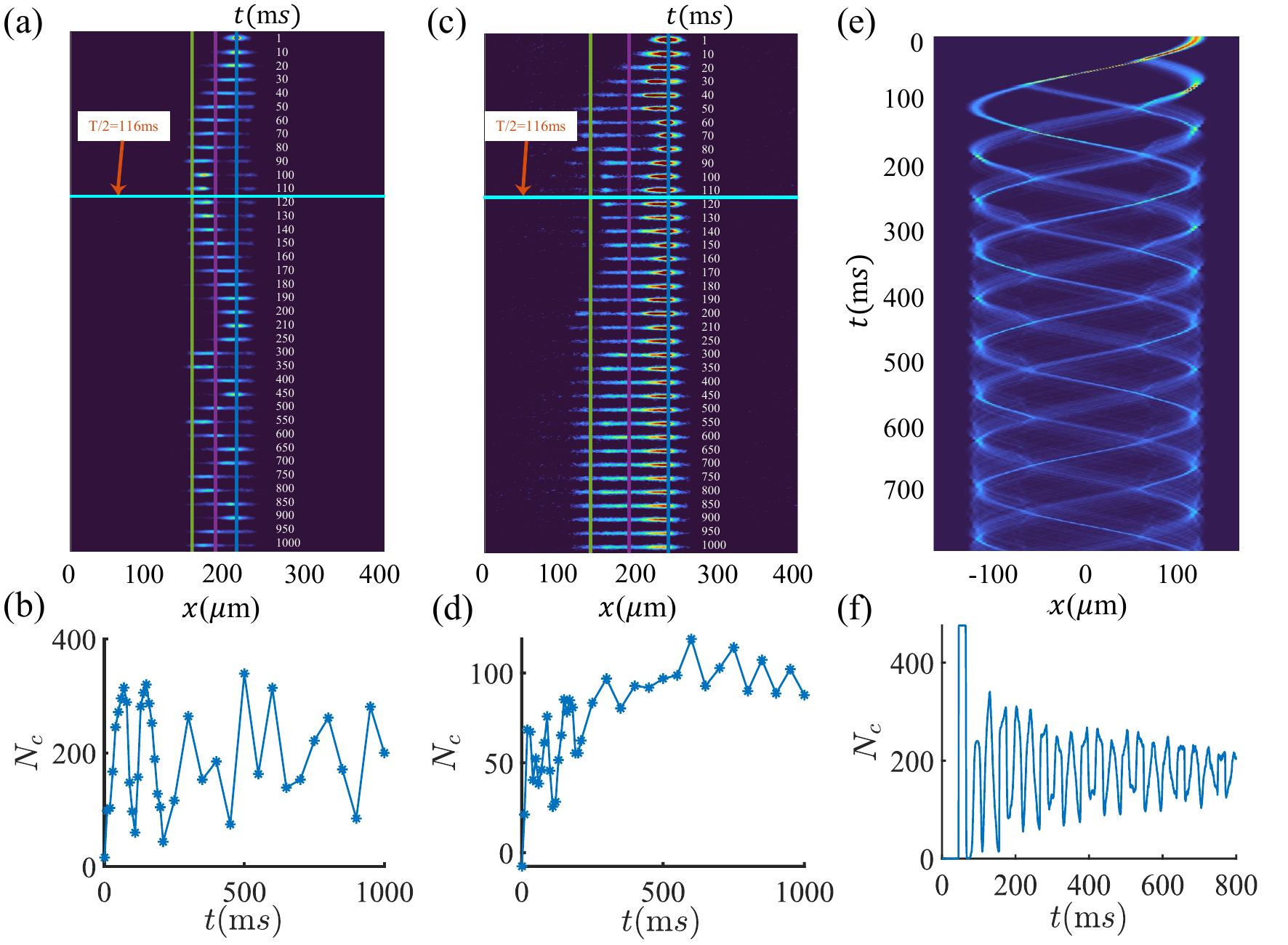}
    \caption{Evolution in the band structure. (a) Evolution of atoms which are initially prepared at a displacement $x_0=70\,\mathrm{\mu m}$ from the trap center. The blue vertical line is the initial location of atoms. The purple vertical line is the trap center. The green vertical line is the symmetrical location to the center of trap. The gradient is sufficiently small that the atoms cannot reach the edge of the Brillouin zone and undergo regular harmonic oscillations. The light blue horizontal line is the half period of the regular oscillation. (b) The evolution of the atom number $N_c$ within a $20\,\mathrm{\mu m}$ region around the trap center, which maintains the oscillation. (c) Evolution of atoms which are initially prepared at a position displaced by $x_0=120\,\mathrm{\mu m}$ from the trap center. The gradient is large enough for atoms to reach the edge of the Brillouin zone, leading to a fast dephasing. (d) The oscillation of $N_c$ becomes faster and rapidly decays to a finite nonzero value. (e) and (f): Simulation of the atomic dynamics within the band structure when $x_0=120\,\mathrm{\mu m}$. The dephasing is almost developed within 4 periods and the oscillation of $N_c$ also rapidly decays.}
    \label{FigS2:exp_dephasing}
\end{figure}

In our experiment, we find that the rapid dephasing of the atomic ensemble is intrinsically linked to Bloch oscillations experienced by atoms located on the slope of the trap; see Fig.~\ref{FigS2:exp_dephasing}(a)-(d). From the time evolution of the optical-density profiles, we observe that dephasing is extremely slow in Fig.~\ref{FigS2:exp_dephasing}(a) when the atoms undergo regular harmonic oscillations. After around four periods, the ensemble remains almost fully coherent. In contrast, when a fraction of the atoms becomes localized on the slope in Fig.~\ref{FigS2:exp_dephasing}(c), the atoms that slide down dephase completely within the same interval. Quantitatively, we evaluate the atom number $N_c$ within a $20\,\mathrm{\mu m}$ region around the trap center. In the absence of dephasing, $N_c$ oscillates; see Fig.~\ref{FigS2:exp_dephasing}(b). Once dephasing occurs, the oscillation amplitude decays. After complete dephasing $N_c$ approaches to a steady value, see Fig.~\ref{FigS2:exp_dephasing}(d).

This phenomenon can be explained as follows. In a harmonic potential dressed by a shallow lattice, atoms feel a potential gradient and therefore evolve within the corresponding band structure. For a fixed quasi-1D confinement formed by the 1560-nm beam, we tune the magnitude of this gradient by varying the initial displacement $x_0$ of the BEC from the trap center. When $x_0$ is small, the gradient is insufficient to drive the atoms to the edge of Brillouin-zone, and the atoms simply oscillate in the trap. For larger $x_0$, however, the atoms reach the Brillouin-zone edge at $x_1$ during their downward motion. Owing to the extremely shallow lattice, the dynamics then split into two pathways: with some probability the atoms remain in the first band and undergo Bloch oscillations, while with complementary probability they undergo Landau–Zener tunneling into the second band and continue their downward motion. Within the experimental parameter range, the potential energy of the atoms is not large enough to drive them to ascend the third band. Therefore, we only consider two bands. When the atoms pass through the bottom of the trap and reach the symmetric position on the opposite side of the trap at $-x_1$, they return to the edge of the Brillouin-zone. A similar process occurs: they may tunnel back into the first band and climb further up the potential, or remain in the second band and be ``reflected'' back. This reflection takes place earlier than half a harmonic period, introducing an additional phase dispersion. This band-mediated reflection mechanism is responsible for the rapid dephasing observed in the ensemble. 

We demonstrate this process by varying the displacement in a fixed trap. The trap frequency is $\omega_{\parallel}=2\pi\times 4.3\text{Hz}$, smaller than the one used in the main text. This is because the gradient of the trap in the main text is so large that the ``oscillation'' displacement needs to be very small, which is not conducive to experimental observation. So we choose $\omega_{\parallel}=2\pi\times 4.3\text{Hz}$, where ``oscillation'' displacement $x_0=70\mathrm{\mu m}$ and ``localization'' displacement $x_0=120\mathrm{\mu m}$; see Fig.~\ref{FigS2:exp_dephasing}(a)-(d).

We also simulate the atomic dynamics within the band structure using an ensemble of classical noninteracting particles. Specifically, 1000 particles are initially distributed around $x_0$ with zero initial velocities. Then they evolve within the band structure. Each time a given particle reaches the Brillouin-zone edge, it undergoes a Landau–Zener transition with a probability $P_{LZ}$ \cite{Landau1932EnergyTransferII,Zener1932NonAdiabatic}. This probability can be calculated from the relevant trap parameters:
\begin{equation}
    P_{LZ}=\mathrm{exp}(-\frac{2\pi\Delta^2}{\hbar|\beta|})
\end{equation}
where $\Delta$ is the half energy gap between the first and second bands, 
\begin{equation}
    \beta=\left.\frac{\mathrm{d}}{\mathrm{d}t}(E_1-E_2) \right |_{crossing}=\left. \frac{F}{\hbar}\frac{\partial}{\partial k}(E_1-E_2)\right |_{k=G/2}
\end{equation}
is the sweeping rate of the energy difference near crossing point under the two-level basis of plane-wave states $\left | k  \right \rangle $ and $\left | k-G  \right \rangle $, $F$ is the force provided by the gradient of the trapping potential. For shallow lattice, the dispersion near the BZ boundary $k=G/2(G=2\pi/a)$ can be approximated by the free-particle form $E_k^0=\hbar^2k^2/(2m)$, yielding
\begin{equation}
    \left.\partial_k(E_1-E_2)\right |_{k=G/2}=\frac{\hbar^2G}{m}.
\end{equation}

Once tunneling is taken into account, the trajectories of all particles are computed. Now we have the distribution of the ensemble at any given time; see Fig.~\ref{FigS2:exp_dephasing}(e). We can observe that some atoms are reflected back before half a period of regular harmonic oscillation. This phenomenon leads to a fast dephasing process. The dephasing is nearly complete within about four periods. Atom number $N_c$ within a $20\,\mathrm{\mu m}$ region around the trap center is also calculated in Fig.~\ref{FigS2:exp_dephasing}(f). We can observe the fast decay of oscillation.

From both experiment and numerical simulations, we clearly demonstrate that the shallow lattice strongly accelerates the first stage of relaxation, driving the atoms to rapidly relax into an approximately microcanonical ensemble. However, we note that the collision still exists in the first stage, especially when scattering length and initial atomic width are large. A typical phenomenon is that the waist of the narrow ring $\sigma_{\rho}$ becomes larger than initial atomic waist $\Delta_x$ after dephasing.

\section{Machine learning algorithm: denoising and parameters extraction}
\subsection{Model Introduction}
To enable a quantitative study of thermalization, we have two tasks. First, we denoise the experimentally measured density profiles. Then, we extract the parameters of the Wigner functions from the denoised distributions. It is a challenging inverse problem, so we employ a deep-learning model to extract the underlying Wigner functions. Among various architectures, the denoising autoencoder (DAE) is particularly well suited for our purpose \cite{vincent2010stacked}. A DAE performs self-supervised learning by reconstructing clean data from noisy inputs, thereby capturing the essential features of the distribution while suppressing irrelevant fluctuations. The network architecture of DAE is quite simple. The DAE consists of a fully connected symmetric encoder–decoder architecture. The encoder maps the input atomic density profile $n(x)$, discretized into $N$ spatial pixels, to a low-dimensional latent layer. The decoder reconstructs the denoised output from this latent layer.

Specifically in our denoising task, the encoder maps the input into a low-dimensional latent representation through two hidden layers with progressively reduced widths, $$N=400\rightarrow256\rightarrow128.$$ All hidden layers use rectified linear unit (ReLU) activation function. The decoder mirrors the encoder with the inverted structure, $$128\rightarrow256\rightarrow N=400,$$ while the last layer uses a sigmoid activation function. To enable denoising, the training input is a database with noise, whereas the target output remains noise-free. The network is trained by minimizing the mean-squared error (MSE) loss between the clean target distribution and the reconstructed output.

In the parameter extraction task, the encoder structure is the same as in the denoising task, meanwhile the decoder continues to reduce the widths, $$128\rightarrow64\rightarrow32\rightarrow 2~(\rho_0, \sigma_{\rho}).$$ The first two layers use the ReLU activation function, while the last parameter-prediction layer employs a linear layer without any nonlinear activation. The training input is the noise-free database of the dephased profile, whereas the target outputs are the parameters of corresponding Wigner functions. The network is trained by minimizing the mean-squared error (MSE) loss between the target parameters and the predicted output.

\subsection{Synthetic Data Generation and Neural-network Training}
Since the model can only learn the patterns presented in the training database, the coverage of the database plays a crucial role in the data-processing procedure. However, constrained by computational resources and training costs, our dataset cannot exhaustively cover all possible distributions in phase-space. We therefore impose a set of physically motivated assumptions on the Wigner functions. We assume the Wigner functions follow the ansatz
\begin{equation}
f(x,\tilde{p}) = A \exp\left\{-\frac{(\rho-\rho_0)^2}{2\sigma_{\rho}^2}\right\}, \label{wigner_uniform}
\end{equation}
where $\rho^2 = x^2 + \tilde{p}^{2}$, $\tilde{p}=p/(m\omega_{\parallel})$ is the normalized momentum along $\hat{x}$. The angular distribution is taken to be uniform. And since the initial displacement $x_0=120\mathrm{\mu m}$, the value of $\rho$ and $\sigma_{\rho}$ are restricted within a certain range. Based on these assumptions, the specific parameter values are summarized in Table.~\ref{table_paramter_post_dephasing}.
\begin{table}[!ht]
    \centering
    \caption{Parameter value of post-dephasing database}
    \begin{tabular}{|c|c|c|}
    \hline
        Parameter               & Value range       & Number of Samples \\ \hline
        $\rho_0(\mathrm{\mu m})$         & $0\sim225$      & 301 \\ \hline
        $\sigma_{\rho}(\mathrm{\mu m})$  & $1\sim100$        & 196 \\ \hline
        \multicolumn{2}{|c|}{Total number of samples} & 58996 \\ \hline
    \end{tabular}
    \label{table_paramter_post_dephasing}
\end{table}

At this stage, we have generated 58996 Wigner functions, sufficient to cover the distributions expected in the experiment. From each Wigner function, we compute a smooth atomic density profile $n(x)$. Then we need to add some noise to these profiles. We acquired 2000 background images without atoms in the experiment. For each density sample, 25 images of noise are randomly selected and averaged to produce the background noise, yielding 58996 noise data. To further account for variations in noise amplitude, we scale the noise to create 10 different amplitude levels. The scaled noise $\delta n(x)$ is then added directly to the smooth density profiles, resulting in a final dataset of 589960 noisy samples that are used to train the denoising task. For the denoising task, we take $n(x)+\delta n(x)$ as input and $n(x)$ as target to train the denoising model $\mathrm{DAE_1}$ \cite{paszke2019pytorch}. For parameter extraction task, we take $n(x)$ as input and $(\rho_0, \sigma_{\rho})$ as target to train the parameter extraction model $\mathrm{DAE_2}$. After training, we apply $\mathrm{DAE_1}$ to the experimental data $n_{exp}$, resulting in the denoised density $n_{pure}$. Then we apply $\mathrm{DAE_2}$ to the pure data $n_{pure}$, resulting in the parameters $(\rho_0, \sigma_{\rho})$ of the Wigner functions.

\begin{figure}[!ht]
    \centering
    \includegraphics[width=0.65\linewidth]{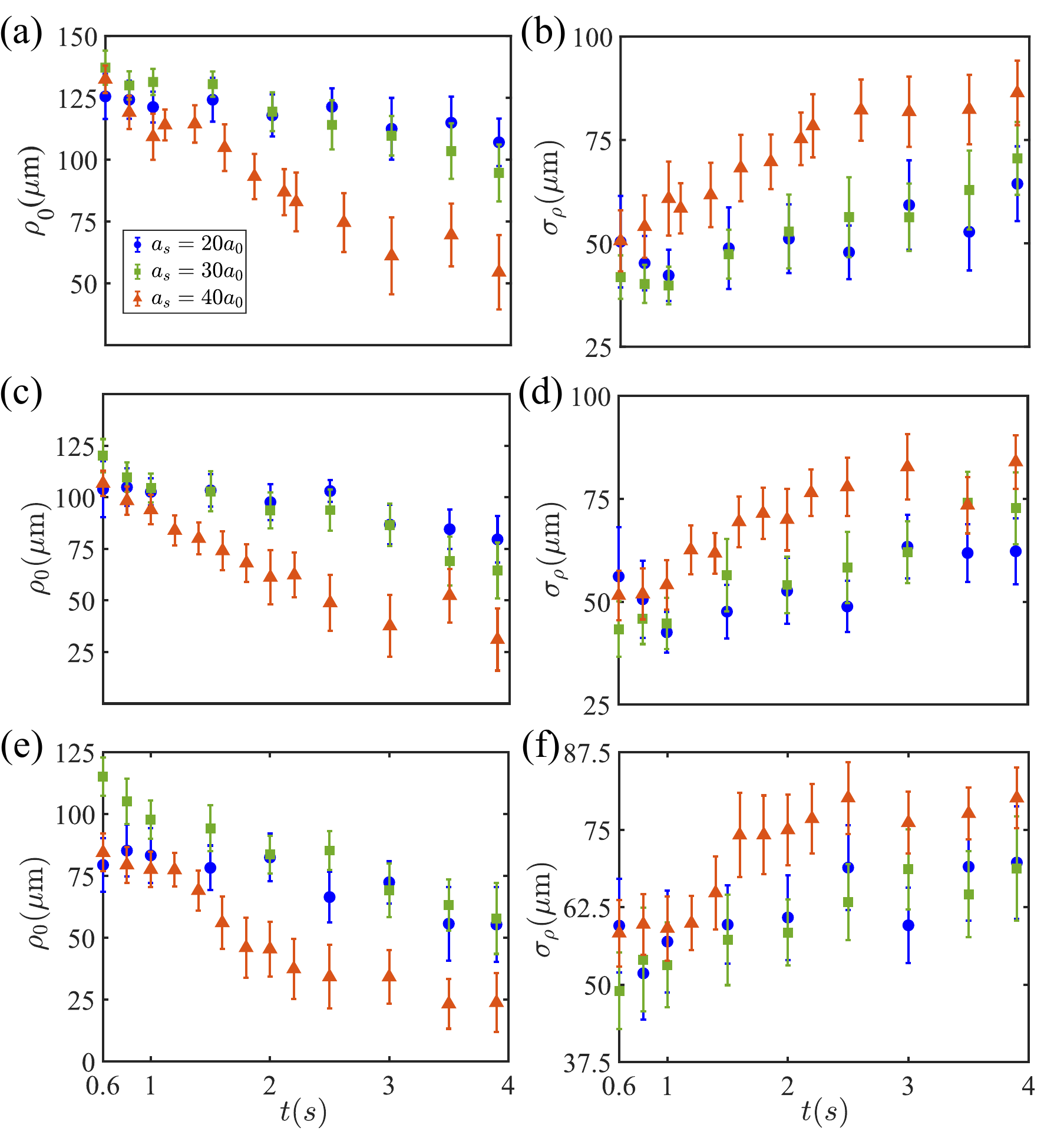}
    \caption{The time evolution of $\rho_0$ (panels a, c, e) and $\sigma_\rho$ (panels b, d, f) is shown for three interaction strengths, $20a_0$ (blue circles), $30a_0$ (green square) and $40a_0$ (orange triangles), for initial spatial widths of BEC $\Delta_x=7\mathrm{\mu m}$ (a, b), $\Delta_x = 10\mathrm{\mu m}$ (c, d) and $\Delta_x = 13.6\mathrm{\mu m}$ (e, f) respectively. Here $a_0$ is the Bohr radius. }
    \label{Fig:all_data}
\end{figure}

Using these models we successfully recover smooth density profiles and the parameters of the Wigner functions from the noisy experimental profiles. The results are shown in Fig.~\ref{Fig:all_data}. The reliability of the extraction is demonstrated in the main text.

\section{Modified Boltzmann equation}
It is well known that at finite temperature, particles will thermalize to the canonical ensemble. However, our experiment indicates that the thermalization process is exceptionally slow. The physical origin is that a strictly one-dimensional Bose gas is close to an integrable system, for which elastic two-body collisions do not redistribute the axial single-particle energy. In the present quasi-one-dimensional setting, this constraint is weakly relaxed by the residual coupling to transverse degrees of freedom. Motivated by this picture, we introduce a modified Boltzmann equation that incorporates a small stochastic mismatch of the axial kinetic energy during collisions and use it to simulate the slow relaxation to the canonical ensemble.

To simulate how the system evolves, we use the 1D Boltzmann equation: 
\begin{equation}
    \frac{\partial f}{\partial t}+\frac{p}m\frac{\partial f}{\partial x}+F(x)\frac{\partial f}{\partial p}=I_{\mathrm{coll}}[f],
\end{equation}
where $I_{\mathrm{coll}}[f]$ is the collision term. It comprises two parts:
\begin{equation}
    I_{\mathrm{coll}}[f_1]=\left(\frac{\partial f_1}{\partial t}\right)_{\mathrm{coll}}^{(+)}-\left(\frac{\partial f_1}{\partial t}\right)_{\mathrm{coll}}^{(-)}.
\end{equation}
The leaving term is 
\begin{equation}
    \left(\frac{\partial f_1}{\partial t}\right)_{\mathrm{coll}}^{(-)}=\int \D p_2\frac{|p_1-p_2|}{m}f_1(x,p_1,t)f_1(x,p_2,t),
\end{equation}
which describes the process in which two particles leave $\mathrm{d}p_1\mathrm{d}p_2$ after collision. And the entering term is
\begin{equation}
    \left(\frac{\partial f_1}{\partial t}\right)_{\mathrm{coll}}^{(+)}\D p_1 =\int\D p'_1 \D p'_2\frac{|p'_1-p'_2|}{m}f_1(x,p'_1,t)f_1(x,p'_2,t){\rm T}(p_1;p'_1,p'_2),
    \label{entering1}
\end{equation}
where the ${\rm T}$ matrix vanishes unless collision between $p'_1$ and $p'_2$ results in two particles in which one carries momentum $p_1$.

If we consider elastic collision $|p_1-p_2|=|p'_1-p'_2|$, then the Jacobian $\left|\frac{\p (p_1,p_2)}{\p (p'_1,p'_2)}\right|=1$, which gives us
\begin{equation}
    \D p_1\D p_2=\D p'_1\D p'_2.
\end{equation}
Using this relation, the collision integral can be written as
\begin{equation}
    I_\mathrm{coll}[f_1]=\int \D p_2\frac{|p_1-p_2|}{m}(f_1^{\prime}f_{2}^{\prime}-f_1f_{2}),
\end{equation}
where $f_1=f(x,p_1,t),f_2=f(x,p_2,t),f'_1=f(x,p'_1,t),f'_2=f(x,p'_2,t)$. In one dimension, elastic two-body collisions lead to a simple exchange of momenta between the colliding particles, $(p'_1,p'_2)=(p_2,p_1)$. As a consequence, the detailed balance condition is automatically satisfied, yielding, $f_1^{\prime}f_{2}^{\prime}-f_1f_{2}=0$. Because the collision term is strictly zero, all particles merely evolve along constant-energy contours in phase space, and the overall distribution never changes if $f$ depends only on energy. In order to simulate the effects of losing or gaining the kinetic energy belonging to the axial dimension, we consider inelastic collision. We assume that the total axial kinetic energy changes by an amount $\epsilon$ during each collision, governed by a probability density $g(\epsilon)$.

In entering collision term, the pre- and post-collision momenta are $(p'_1,p'_2)$ and $(p_1,p_2)$, which satisfy
\begin{equation}
    \begin{gathered}
        p'_1+p'_2=p_1+p_2,\\
        \frac{1}{2m}{p'_1}^2+\frac{1}{2m}{p'_2}^2+\epsilon=\frac{1}{2m}p_1^2+\frac{1}{2m}p_2^2.
    \end{gathered}\label{entering}
\end{equation}
The conditions for a collision to occur are $\Delta_+=(p_1-p_2)^2-4m\epsilon\geq 0$, and the solution of $p'_1,p'_2$ is
\begin{equation}
    p'_1=\frac{p_1+p_2}{2}+\frac{\sqrt{(p_1-p_2)^2-4m\epsilon}}{2},\quad p'_2=\frac{p_1+p_2}{2}-\frac{\sqrt{(p_1-p_2)^2-4m\epsilon}}{2}.
\end{equation}
Using Eq.~(\ref{entering}), we obtain
\begin{equation}
\begin{gathered}
|p'_1-p'_2|\D p'_1\D p'_2=|p_1-p_2|\D p_1\D p_2.\label{diff_rel}
\end{gathered}
\end{equation}
The entering collision term is
\begin{equation*}
    \left(\frac{\partial f_1}{\partial t}\right)_{\mathrm{coll}}^{(+)}\D p_1=\int \D\epsilon\int_{\Delta_+\geq 0} \D p'_1\D p'_2~g(\epsilon)\frac{\left|p'_1-p'_2\right|}{m}f(x,p'_1(p_1,p_2,\epsilon),t)f(x,p'_2(p_1,p_2,\epsilon),t).
\end{equation*}
Using Eq.~(\ref{diff_rel}), we have
\begin{equation}
    \left(\frac{\partial f_1}{\partial t}\right)_{\mathrm{coll}}^{(+)}=\int \D\epsilon\int_{\Delta_+\geq 0} \D p_2~g(\epsilon)\frac{\left|p_1-p_2\right|}{m}f(x,p'_1(p_1,p_2,\epsilon),t)f(x,p'_2(p_1,p_2,\epsilon),t).
\end{equation}
For the leaving collision term, the pre- and post-collision momenta are $(p_1,p_2)$ and $(p'_1,p'_2)$, so the conditions for a collision to occur are $\Delta_-=(p_1-p_2)^2+4m\epsilon\geq 0$. The leaving collision term is
\begin{equation}
    \left(\frac{\partial f_1}{\partial t}\right)_{\mathrm{coll}}^{(-)}=\int \D\epsilon\int_{\Delta_-\geq 0} \D p_2~g(\epsilon)\frac{\left|p_1-p_2\right|}{m}f(x,p_1,t)f(x,p_2,t).
\end{equation}
The collision term can be combined as
\begin{equation}
    I_{\mathrm{coll}}[f_1]=\int \D\epsilon\int\D p_2~g(\epsilon)\frac{\left|p_1-p_2\right|}{m}\left[f'_1f'_2\Theta(\Delta(p_1,p_2,\epsilon))-f_1f_2\Theta(\Delta(p_1,p_2,-\epsilon))\right],
\end{equation}
where $\Theta$ is the step function and $\Delta(p_1,p_2,\epsilon)=(p_1-p_2)^2-4m\epsilon$. The equilibrium condition of our Boltzmann equation is
\begin{equation}
    \int \D\epsilon~g(\epsilon)\left[f(x,p'_1,t)f(x,p'_2,t)\Theta(\Delta(p_1,p_2,\epsilon))-f(x,p_1,t)f(x,p_2,t)\Theta(\Delta(p_1,p_2,-\epsilon))\right]=0.
\end{equation}
For the system to reach a stationary thermal state, this condition must be satisfied. We expect the Maxwell-Boltzmann distribution to satisfy the equilibrium condition of the Boltzmann equation, which means
\begin{equation}
    \begin{gathered}
        \begin{aligned}
            \int \D\epsilon~g(\epsilon)\bigg[ & f^{\text{(MB)}}(x,p'_1)f^{\text{(MB)}}(x,p'_2)\Theta(\Delta(p_1,p_2,\epsilon)) \\
            & - f^{\text{(MB)}}(x,p_1)f^{\text{(MB)}}(x,p_2)\Theta(\Delta(p_1,p_2,-\epsilon)) \bigg] = 0,
        \end{aligned}\\[2ex]
        f^{\text{(MB)}}(x,p)=C\exp{\left[-\beta\left(\frac{p^2}{2m}+V(x)\right)\right]}.
    \end{gathered}  
\end{equation}
For the Maxwell--Boltzmann distribution,
\begin{equation}
f^{\mathrm{(MB)}}(x,p'_1)f^{\mathrm{(MB)}}(x,p'_2)
=
\exp(\beta \epsilon)\,
f^{\mathrm{(MB)}}(x,p_1)f^{\mathrm{(MB)}}(x,p_2),
\end{equation}
because Eq.~(\ref{entering}) implies that the total axial energy of the incoming pair is smaller than that of the outgoing pair by $\epsilon$. Substituting this relation into the equilibrium condition yields
\begin{equation}
\int_{-\infty}^{\infty}\D\epsilon~g(\epsilon)\left(\E^{\beta\epsilon}\Theta(\Delta(p_1,p_2,\epsilon))-\Theta(\Delta(p_1,p_2,-\epsilon))\right)=0.
\end{equation}
In our experimental regime, the energy change $\epsilon$ is much smaller than the typical axial kinetic energy, so the kinematic constraints imposed by the step functions can be neglected to leading order. The equilibrium condition then reduces to
\begin{equation}
    \int_{-\infty}^{\infty}\D\epsilon~g(\epsilon)\left(\E^{\beta\epsilon}-1\right)=0.\label{Equi_cond}
\end{equation}
This condition is satisfied if $g(\epsilon)$ is the product of $\E^{-\beta\epsilon/2}$ and an even function,
\begin{equation}
    g(\epsilon)=\E^{-\beta\epsilon/2}h(\epsilon),
\end{equation}
and the even function $h(\epsilon)$ should decay fast enough to ensure the normalization condition $\int g(\epsilon)\D\epsilon=1$. We can write down some specific forms of $h(\epsilon)$ satisfying these conditions. For example,

(i) the Gaussian ansatz:
\begin{equation}
    h(\epsilon) \propto \exp\left[-\frac{\epsilon^2}{2\sigma^2}\right],
\end{equation}

(ii) the Lorentzian-Gaussian ansatz (with $w = 1.38\sigma$):
\begin{equation}
    h(\epsilon) \propto \frac{1}{\epsilon^2 + w^2} \exp\left[-\frac{\epsilon^2}{2w^2}\right],
\end{equation}

(iii) the step-function-like ansatz:
\begin{equation}
    h(\epsilon) \propto \Theta(\sqrt{3}\sigma - |\epsilon|).
\end{equation}
Because $\beta\sigma\ll 1$ in our experimental regime, the standard deviation $\sigma$ of the probability distribution $h(\epsilon)$ is approximately equal to that of $g(\epsilon)$. This characteristic energy uncertainty, $\sigma$, can be independently estimated from the two-body scattering theory in a harmonic waveguide, as we will show in the next section. We have two parameters $\beta$ and $\sigma$ in this phenomenological model. $\beta$ sets the temperature of the final canonical ensemble, and $\sigma$ represents the typical axial energy mismatch during a collision, thus controlling the rate of the thermalization. Furthermore, we find that the specific form of $h(\epsilon)$ does not significantly affect the thermalization dynamics, as long as it satisfies the equilibrium condition Eq.(\ref{Equi_cond}) and has the same standard deviation $\sigma$.

\section{Estimation of the energy mismatch}
To justify the phenomenological energy mismatch distribution $h(\epsilon)$ used in our modified one-dimensional Boltzmann equation and to independently estimate the dimensionless parameter $\eta$, we utilize the multi-channel two-body scattering theory in a harmonic waveguide \cite{Moore2004TightWaveguides}.

Consider two interacting atoms of mass $m$ confined in a waveguide with a transverse harmonic trapping frequency $\omega_\perp$. The two-body problem is separable, and the relative motion is governed by the reduced mass $\mu = m/2$. We define the relative transverse harmonic oscillator length as $a_{\perp, \mathrm{rel}} = \sqrt{\hbar/(\mu\omega_\perp)} = \sqrt{2}a_\perp$, where $a_\perp = \sqrt{\hbar/(m\omega_\perp)}$ is the single-particle oscillator length defined in the main text.

During a collision, the relative motion state, initially in the transverse mode $n$ with an axial wavevector $k_n$, is scattered into a superposition of available transverse modes $|n', m_l=0\rangle$. The asymptotic scattered wavefunction can be expressed as:
\begin{equation}
\langle \rho\phi z | \psi(\mathcal{E}) \rangle = \sum_{n'=0}^\infty \langle \rho\phi | n'0 \rangle \left[ \delta_{n',n} e^{ik_n z} + f(k_{n'} \leftarrow k_n)_{n' \leftarrow n} e^{i k_{n'} |z|} \right],
\end{equation}
where $(\rho, \phi, z)$ denote the relative cylindrical coordinates, and $\mathcal{E}=E/(2\hbar\omega_\perp)-1/2=(a_{\perp, \mathrm{rel}} k_{n}/2)^2+n$ is the scaled conserved total energy. The inter-band scattering amplitude $f(k_{n'} \leftarrow k_n)_{n' \leftarrow n}$ is analytically derived as a function of the 3D s-wave scattering length $a_\mathrm{s}$:
\begin{equation}
f(k_{n'} \leftarrow k_n)_{n' \leftarrow n} = -\frac{2i}{a_{\perp, \mathrm{rel}} k_{n'}} \frac{1}{\frac{a_{\perp, \mathrm{rel}}}{a_\mathrm{s}} + \zeta\left(1/2, -\left(\frac{a_{\perp, \mathrm{rel}} k_n}{2}\right)^2 - n\right)},
\end{equation}
with $\zeta$ being the Hurwitz zeta function. The outgoing axial wavevector $k_{n'}$ for channel $n'$ is determined by energy conservation: $k_{n'} = \frac{2}{a_{\perp, \mathrm{rel}}} \sqrt{\left(\frac{a_{\perp, \mathrm{rel}} k_n}{2}\right)^2 + n - n'}$. 

To calculate the statistical variance of the energy change per collision, we must determine the transition probability $p_{n'}$ from the initial state $n$ to an available final state $n'$. Based on the conservation of probability current, the total transition probability (summing over both transmitted and reflected waves) is given by $p_{n'} = 2 \frac{k_{n'}}{k_n} |f(k_{n'} \leftarrow k_n)_{n' \leftarrow n}|^2(n' \neq n)$. Since total energy is conserved during the collision, the axial energy change for a transition into channel $n'$ is $\Delta E_{n' \leftarrow n} = (n - n')\hbar\omega_\perp$.

\begin{figure}[htbp]
    \centering
    \includegraphics[width=0.5\textwidth]{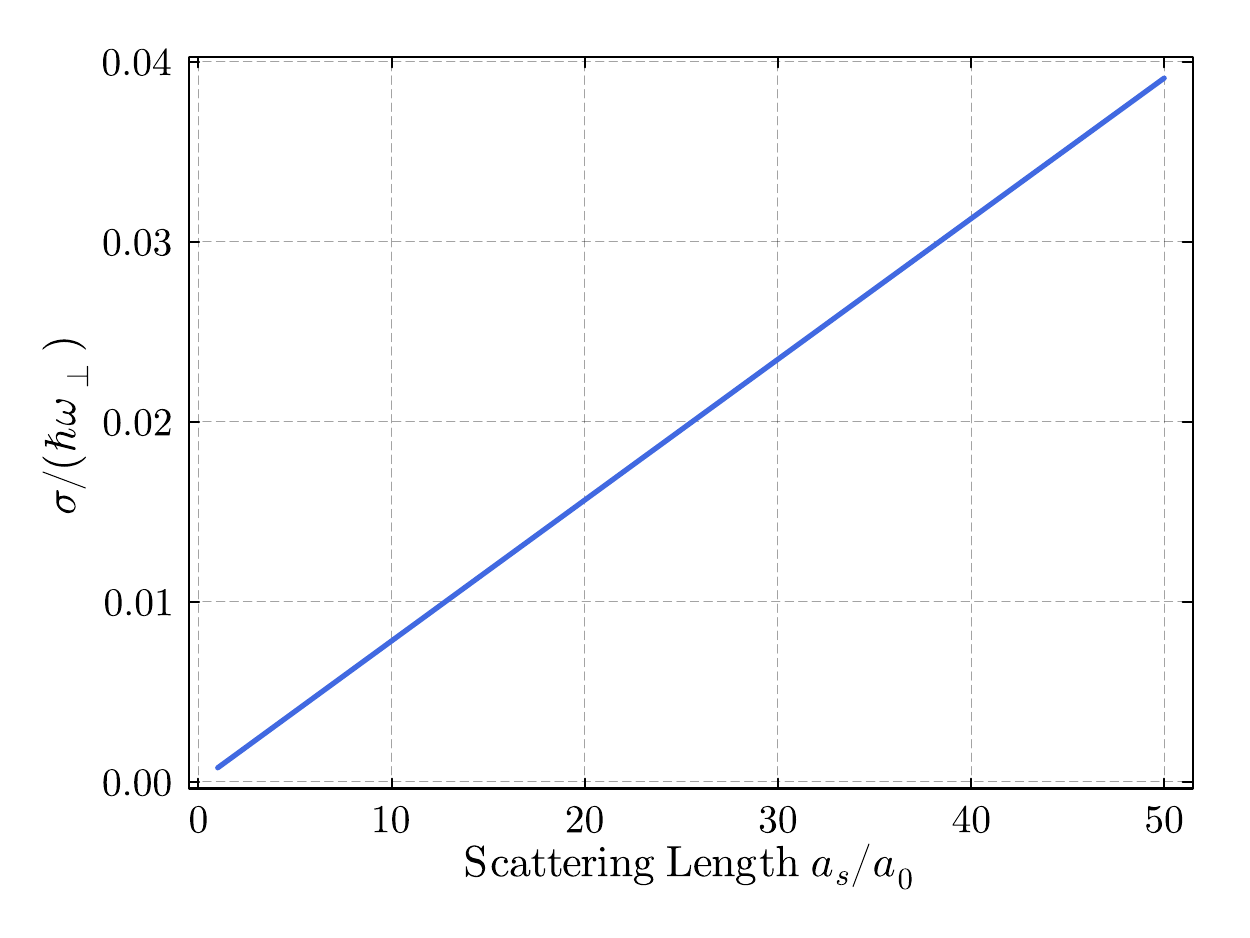}
    \caption{Numerical calculation of the energy uncertainty $\sigma$ as a function of the 3D scattering length $a_\mathrm{s}$. The linear fit yields the dimensionless constant $\eta \approx 0.567$ in the relation $\sigma = 4\pi\eta \frac{\hbar^2 a_\mathrm{s}}{m a_\perp^3}$.}
    \label{fig:scatter}
\end{figure}

We evaluate the energy uncertainty $\sigma$, defined as the standard deviation of the axial energy change across all possible scattering channels for a given initial state:
\begin{equation}
\sigma^2(\Delta E_{\mathrm{initial}=n}) = \sum_{n'} p_{n'} \left(\Delta E_{n'} - \langle \Delta E_{\mathrm{initial}=n} \rangle\right)^2.
\end{equation}
For the numerical estimation, we utilize a typical initial kinetic energy characteristic of the microcanonical-like distribution observed in our experiment to determine $k_n$, and we choose $n=0$. In the weakly interacting regime ($a_\mathrm{s} \ll a_\perp$), the transition probabilities $p_{n'}$ scale as $|f|^2 \propto a_\mathrm{s}^2$. We calculate the energy uncertainty $\sigma$ numerically across the specific range of 3D scattering lengths $a_\mathrm{s}$ used in our experiment. Because this experimental range falls within the weakly interacting regime, a linear fit to our numerical results is performed. By equating this numerically extracted linear relation to the ansatz used in our main text:
\begin{equation}
\sigma = 4\pi\eta \frac{\hbar^2 a_\mathrm{s}}{m a_\perp^3},
\end{equation}
we robustly extract the dimensionless constant $\eta$. As shown in Fig.~\ref{fig:scatter}, this microscopic calculation yields $\eta \approx 0.567$, which is in reasonable agreement with the $\eta$ extracted from the fit to our experimental relaxation rate, further confirming the validity of our phenomenological model.

\section{Numerical simulation}

Numerically, we discretize the phase space and use \texttt{DifferentialEquations.jl} \cite{rackauckas2017differentialequations} to solve the Boltzmann equation. We employ a four-point central difference scheme to calculate the gradient of the distribution function. The initial state $f(x,p,t=0)$ is set to the Wigner function after dephasing. 

To perform the simulations, we must first determine the key parameters $\beta$ and $\sigma$ corresponding to our experimental conditions. The inverse temperature parameter $\beta = 1/(k_B T)$ is extracted directly from the fully thermalized real-space density profile. After the slow thermalization process, the measured density eventually exhibits a Gaussian shape with a standard deviation of $\sigma_\mathrm{th} \approx 80\,\mathrm{\mu m}$. Assuming thermal equilibrium in the axial harmonic potential $V(x) = \frac{1}{2}m\omega_\parallel^2 x^2$, the spatial density follows the Boltzmann distribution $n(x) \propto \exp[-V(x)/(k_B T)]$. The final temperature can thus be extracted from the Gaussian width as $k_B T = m\omega_\parallel^2 \sigma_\mathrm{th}^2$, which yields $T \approx 116\,\mathrm{nK}$. Consequently, the dimensionless inverse temperature scaled by the transverse frequency is $\tilde{\beta} \equiv \beta\hbar\omega_\perp \approx 0.213$. Combined with the dimensionless energy uncertainty $\tilde{\sigma} \equiv \sigma/\hbar\omega_\perp \sim 0.01 - 0.03$ independently estimated from the scattering theory, we obtain $\beta\sigma = \tilde{\beta}\tilde{\sigma} \sim 0.004 \ll 1$. This specific parameter evaluation further justifies the approximation $\beta\sigma \ll 1$ assumed in our model, thereby confirming that the standard deviation of the probability distribution $g(\epsilon)$ is practically identical to that of the function $h(\epsilon)$.

To characterize the distribution in phase space and set up our initial state, we utilize the Wigner function associated with the one-body density matrix $\rho(x_1,x_2)=\langle x_1|\hat{\rho}|x_2\rangle$:
\begin{equation}
f(x,p)=\frac{N}{\pi\hbar}\int_{-\infty}^{\infty}\D y\;
\rho\!\left(x+\frac{y}{2},x-\frac{y}{2}\right)e^{-i p y/\hbar}.
\label{eq:wigner_def}
\end{equation}
For the numerical implementation, we use dimensionless variables $X=x/l_0$ and $P=p/(m\omega_\parallel l_0)$, where $l_0 = \sqrt{\hbar/(m\omega_\parallel)}$ is the harmonic-oscillator length, $m$ is the atomic mass, and $\omega_\parallel$ is the axial trap frequency. However, we present the final results using the physical coordinates $x\,(\mathrm{\mu m})$ and $v = p/m\,(\mathrm{\mu m \cdot s^{-1}})$.

To construct the dephased initial condition used in the Boltzmann simulation, we begin with a localized Gaussian wave packet centered at $x_0$ with $p_0=0$, whose density profile has spatial width $\sigma_x$, such that the wave function takes the form
\begin{equation}
\psi_{\text{initial}}(x)=\frac1{(2\pi\sigma_x^2)^{1/4}}\exp\!\left[-\frac{(x-x_0)^2}{4\sigma_x^2}\right].
\end{equation}
The Wigner function of this Gaussian wave function remains Gaussian in phase space,
\begin{equation}
f_{\rm pre}(x,p)=\frac{N}{\pi\hbar}\exp\!\left[-\frac{(x-x_0)^2}{2\sigma_x^2}-\frac{2\sigma_x^2 p^2}{\hbar^2}\right].
\label{eq:wigner_pre}
\end{equation}

In our experiment, the initial BEC cloud prepared at $x_0 = 120\,\mathrm{\mu m}$ undergoes rapid dephasing within $600\,\mathrm{ms}$, relaxing into a ring-like microcanonical ensemble. We model this dephased state using the diagonal ensemble:
\begin{equation}
\rho_{\rm dep}(x_1,x_2)=\sum_i \lambda_i\,\psi_i(x_1)\psi_i^*(x_2),
\end{equation}
where $\psi_i$ are the axial harmonic oscillator eigenstates with weights $\lambda_i = |\langle \psi_i | \psi_{\text{initial}} \rangle|^2$. The corresponding Wigner function $f_{\rm dep}(x,p)$ evaluated via Eq.~(\ref{eq:wigner_def}) serves as the initial state $f(x,p,t=0)$ for the Boltzmann equation.

\begin{figure}[H]
    \centering
    \includegraphics[width=\textwidth]{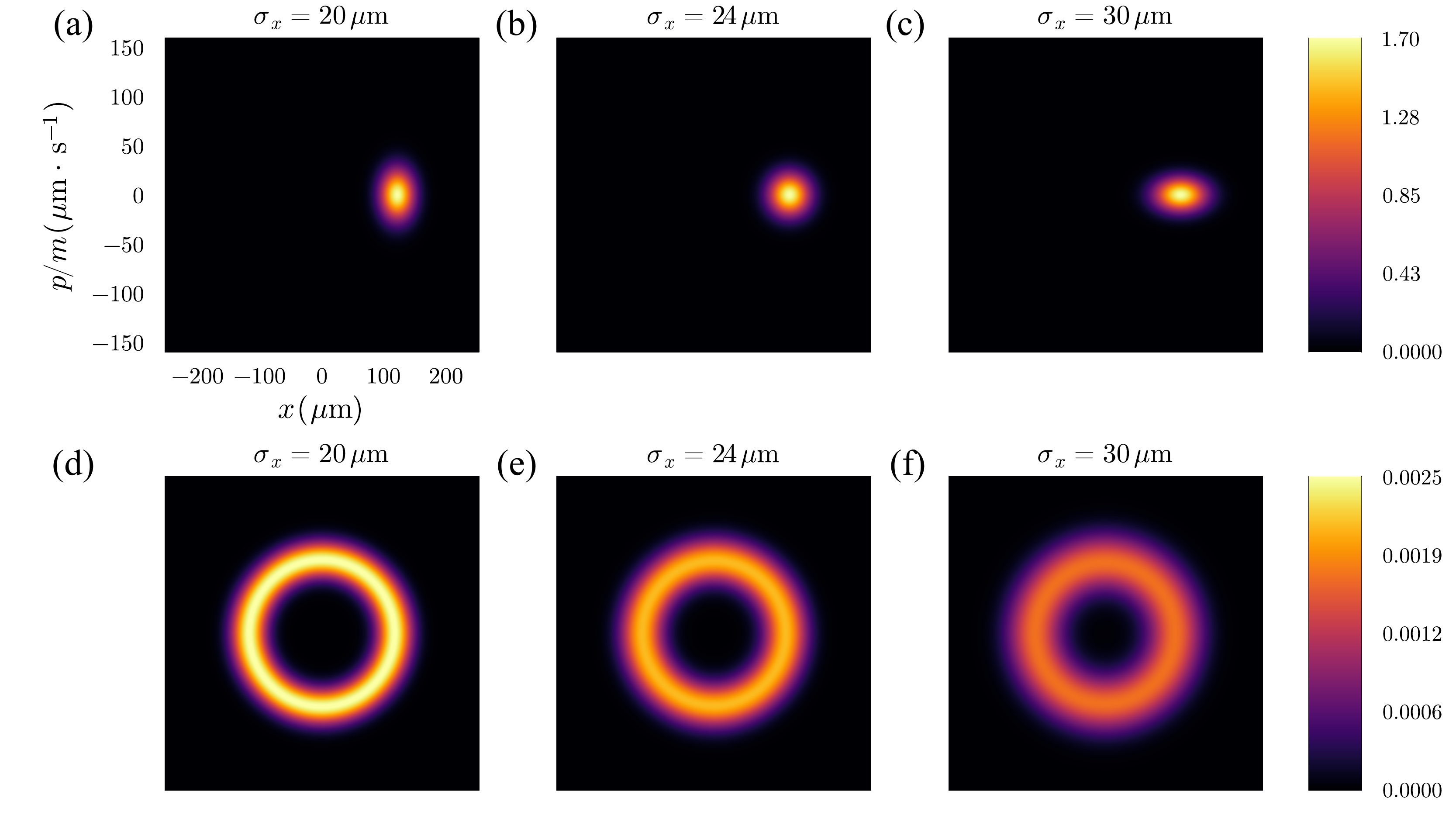}
    \caption{Comparison of Wigner functions for different initial widths. 
    Top row: Wigner function $f_{\text{pre}}(x,p)$ before dephasing for $\sigma_x = 20, 24, 30\,\mathrm{\mu m}$. 
    Bottom row: $f_\text{dep}(x,p)$ after dephasing. The axes are $x\,(\mathrm{ \mu m})$ and $p/m\,(\mathrm{\mu m \cdot s^{-1}})$. After dephasing, $f(x,p)$ is approximately uniform along the constant-energy shell $E(x,p)=p^2/(2m)+V(x)$.}
    \label{fig:wigner_comparison}
\end{figure}

Experimentally, the rapidly dephased state exhibits a characteristic radius-to-width ratio of $\rho/\sigma_\rho \sim 3-4$ in phase space. To faithfully reproduce this condition in our numerical simulations, we set the initial displacement to $x_0 = 120\,\mathrm{\mu m}$ and carefully select the initial spatial width $\sigma_x = 30\,\mathrm{\mu m}$, which yields a dephased Wigner function $f_{\rm dep}$ with $\rho/\sigma_\rho \approx 3.5$, in excellent agreement with the experimental observations. For comparison, Figure~\ref{fig:wigner_comparison} displays the Wigner functions $f_{\rm pre}$ and $f_{\rm dep}$ for initial wavepackets with $\sigma_x=20\,\mathrm{\mu m},24\,\mathrm{\mu m},30\,\mathrm{\mu m}$.

\begin{figure}[H]
    \centering
    \includegraphics[width=\textwidth]{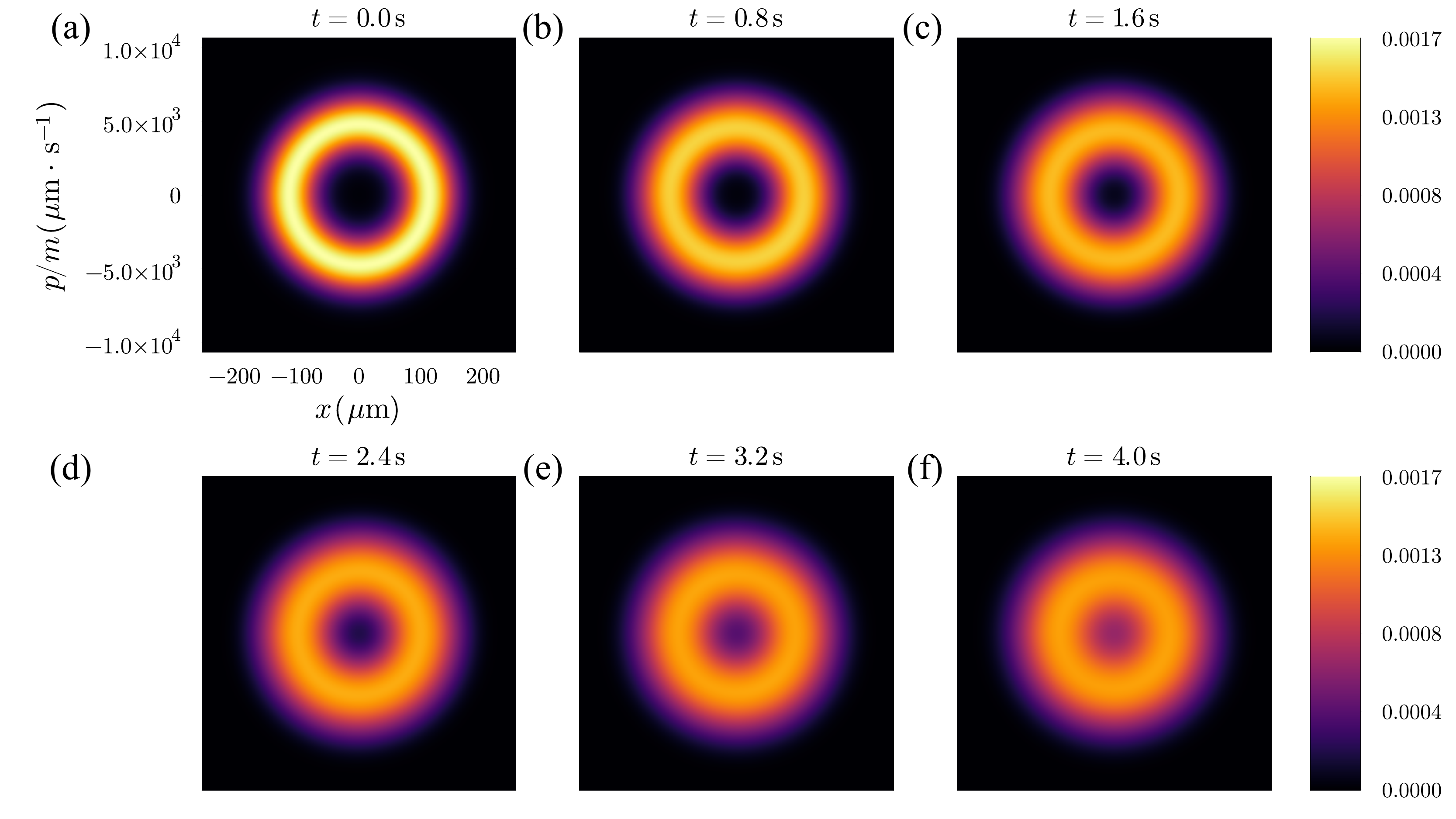}
    \caption{Phase-space evolution of the Wigner function for an energy uncertainty $\sigma/\hbar\omega_\perp = 0.013$ at $t=0, 0.8, 1.6, 2.4, 3.2, 4.0\,\mathrm{s}$. The ring-like energy-shell structure in phase space gradually contracts, indicating slow relaxation. For this parameter set the distribution has not fully thermalized by $t=4~\mathrm{s}$.}
    \label{fig:evolution_sigma_1}
\end{figure}

\begin{figure}[H]
    \centering
    \includegraphics[width=\textwidth]{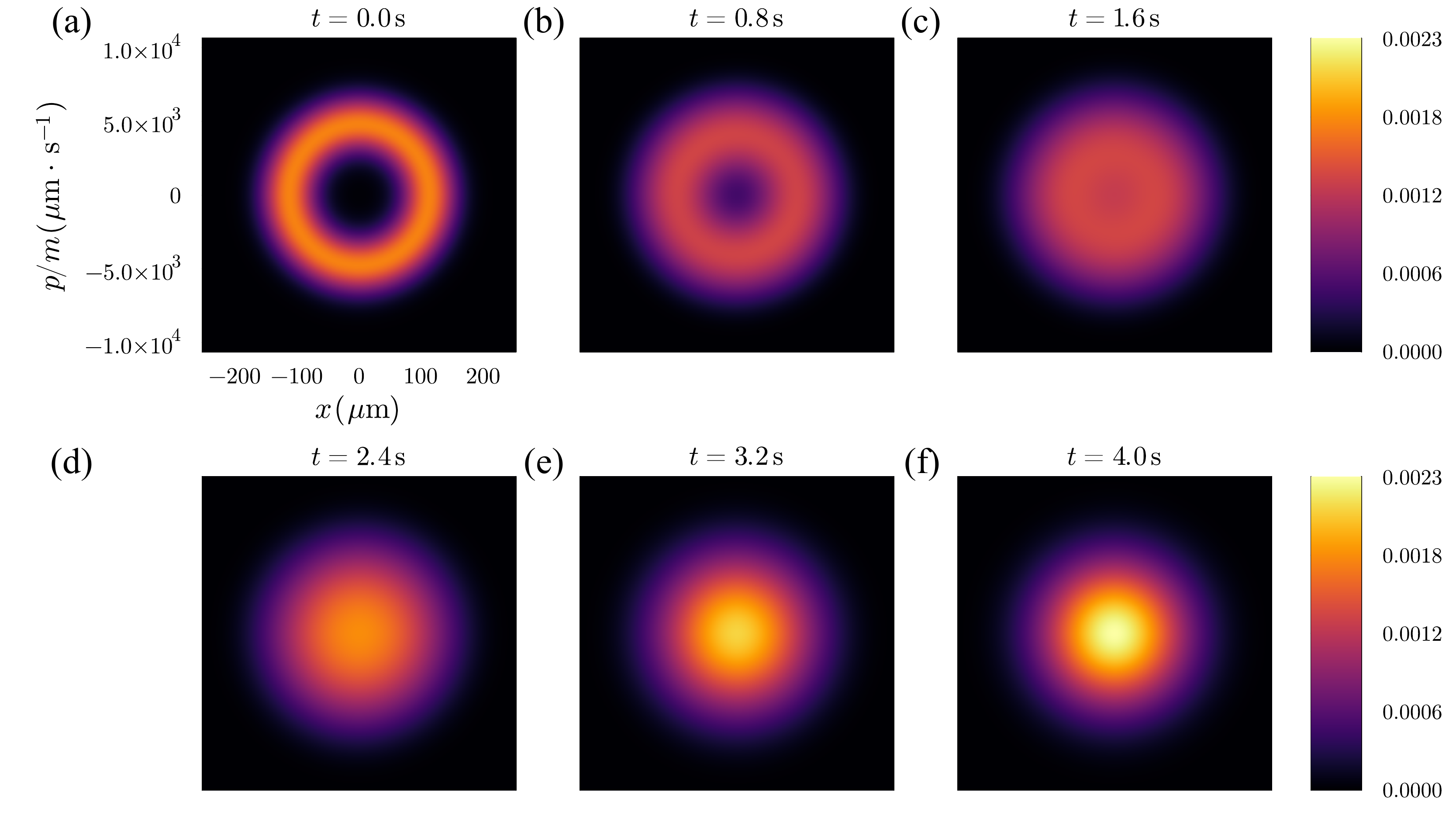}
    
    \caption{Phase-space evolution of the Wigner function for energy uncertainty $\sigma/\hbar\omega_\perp = 0.021$. The plots correspond to the same time sequence as in Fig.~\ref{fig:evolution_sigma_1}. For this parameter set, the system is effectively thermalized by $t=4.0~\mathrm{s}$, and the Wigner function approaches the Maxwell-Boltzmann form.}
    \label{fig:evolution_sigma_2}
\end{figure}

\begin{figure}[H]
    \centering
    \includegraphics[width=\textwidth]{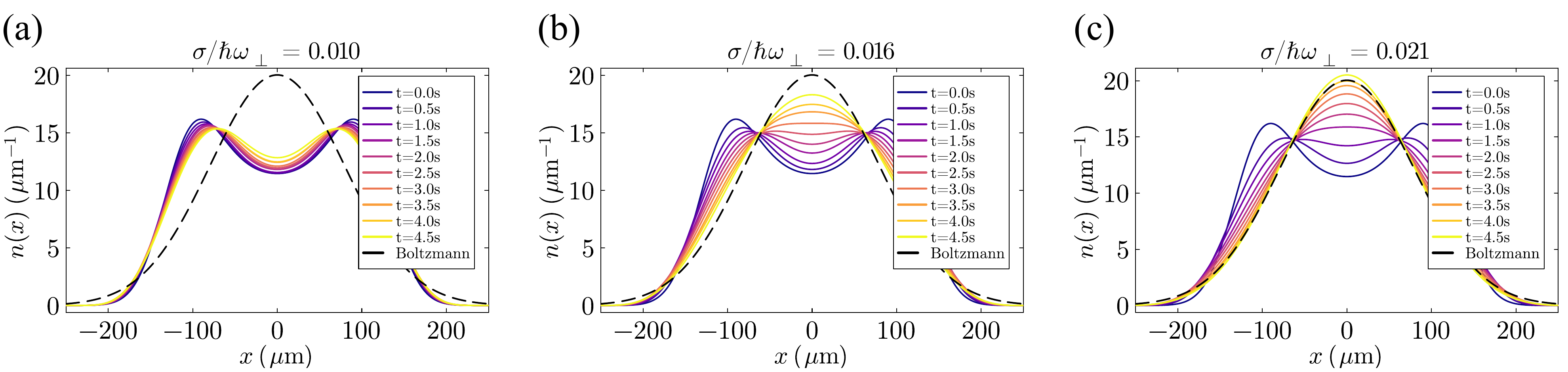}
    \caption{Real-space density evolution $n(x)$ for different energy uncertainties $\sigma/\hbar\omega_\perp = 0.010, 0.016, 0.021$. The different colors represent time evolution from $t=0$ to $4.5\,\mathrm{s}$.}
    \label{fig:th_real_space}
\end{figure}

Figures~\ref{fig:evolution_sigma_1} and \ref{fig:evolution_sigma_2} track the subsequent relaxation driven by inelastic collisions with an energy uncertainty $\sigma$ (in units of $\hbar\omega_\perp$ throughout). For a smaller $\sigma$ (Fig.~\ref{fig:evolution_sigma_1}), the ring-like shell structure contracts slowly and remains distinctly visible at $t=4~\mathrm{s}$, indicating incomplete thermalization within the simulated time. In contrast, for a larger $\sigma$ (Fig.~\ref{fig:evolution_sigma_2}) the Wigner function relaxes faster and approaches a canonical ensemble by $t\approx4~\mathrm{s}$. This trend is consistent with the real-space dynamics in Fig.~\ref{fig:th_real_space}. 

\begin{figure}[H]
    \centering
    \includegraphics[width=\textwidth]{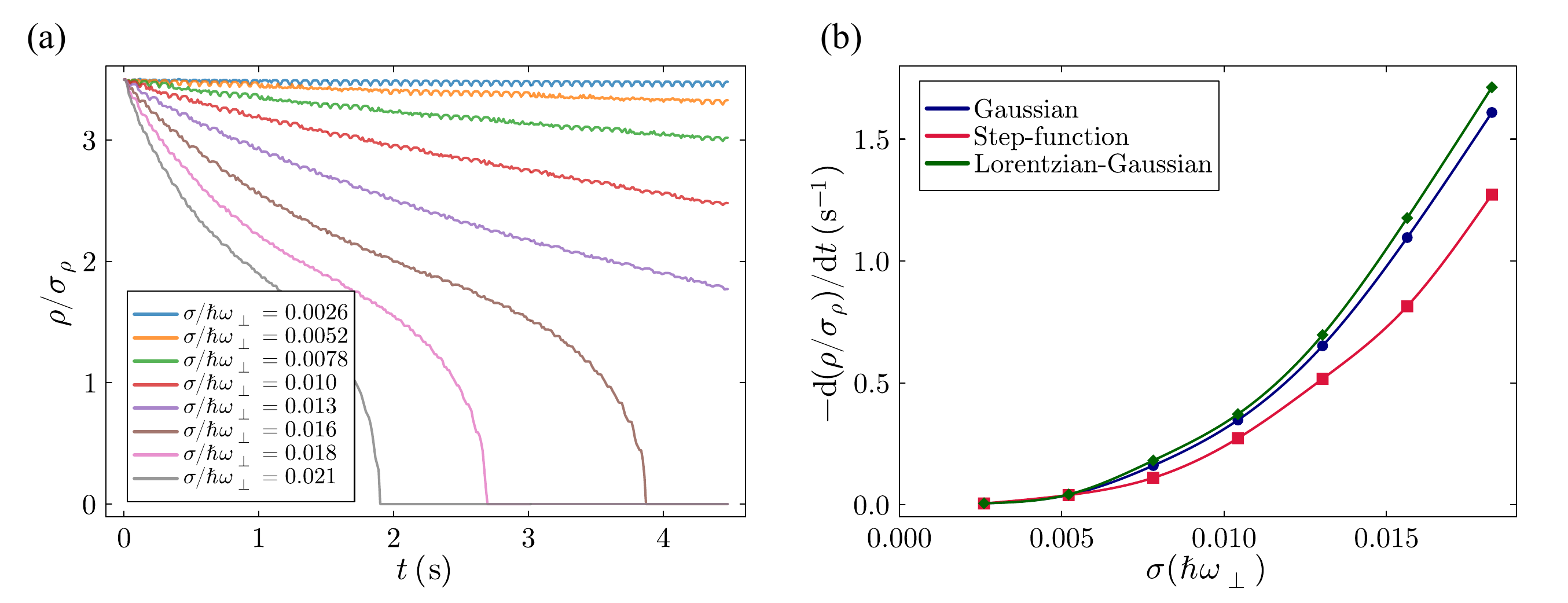}
    \caption{Analysis of the ring-like shell contraction dynamics. 
    (a) Time evolution of $\rho/\sigma_\rho$ for different energy uncertainties $\sigma$, where the color represents the value of $\sigma$. 
    (b) The contraction rate $|\mathrm{d}(\rho/\sigma_\rho)/\mathrm{d}t|$ (fitted from the linear region $t<0.45\,\mathrm{s}$) as a function of $\sigma/\hbar\omega_\perp$.}
    \label{fig:analysis_results}
\end{figure}

To quantify the ring-like structure in phase space, we extract the 1D cross-section of the Wigner function at zero momentum, $f(x, p=0, t)$. Because the dephased Wigner function is symmetric under $x\to -x$, the profile $f(x,p=0,t)$ exhibits two symmetric peaks. For our analysis, we simply isolate the positive half ($x > 0$) of the profile. The shell radius $\rho$ is identified as the position of the density maximum in this region, and the shell width $\sigma_\rho$ is determined by calculating the statistical standard deviation of this isolated half-profile. To quantify the relaxation, Fig.~\ref{fig:analysis_results} extracts the ratio of the shell radius to the shell width $\rho/\sigma_\rho$ and the corresponding contraction rate $|\mathrm{d}(\rho/\sigma_\rho)/\mathrm{d}t|$. The results confirm that a larger $\sigma$ leads to a faster shell contraction, verifying that the energy uncertainty in collisions is the primary factor governing the thermalization rate.    

\newpage

\bibliography{reference}